\documentclass[pdflatex,sn-mathphys-num]{sn-jnl}% Math and Physical Sciences Numbered Reference Style
\usepackage{graphicx}%
\usepackage{multirow}%
\usepackage{amsmath,amssymb,amsfonts}%
\usepackage{amsthm}%
\usepackage{mathrsfs}%
\usepackage[title]{appendix}%
\usepackage{xcolor}%
\usepackage{textcomp}%
\usepackage{manyfoot}%
\usepackage{booktabs}%
\usepackage{algorithm}%
\usepackage{algorithmicx}%
\usepackage{algpseudocode}%
\usepackage{listings}%

\usepackage{bbm}
\usepackage{booktabs}
\usepackage[table]{xcolor}
\usepackage{tabularx}
\definecolor{slateblue}{RGB}{106,90,205}
\usepackage{bbm}
\newcommand{\ind}{\mathbbm{1}}
\usepackage{listings}
\usepackage{placeins}
\theoremstyle{thmstyleone}%
\theoremstyle{thmstyletwo}%

\theoremstyle{thmstylethree}%

\begin{document}

\title[Article Title]{\texttt{PackFlow}: Generative Molecular Crystal Structure Prediction via Reinforcement Learning Alignment}

%%=============================================================%%
%% GivenName	-> \fnm{Joergen W.}
%% Particle	-> \spfx{van der} -> surname prefix
%% FamilyName	-> \sur{Ploeg}
%% Suffix	-> \sfx{IV}
%% \author*[1,2]{\fnm{Joergen W.} \spfx{van der} \sur{Ploeg} 
%%  \sfx{IV}}\email{iauthor@gmail.com}
%%=============================================================%%

\author[1]{\fnm{Akshay} \sur{Subramanian}}\email{akshay\_s@mit.edu}

\author[1]{\fnm{Elton} \sur{Pan}}\email{eltonpan@mit.edu}

\author[1]{\fnm{Juno} \sur{Nam}}\email{junonam@mit.edu}

\author[2]{\fnm{Maurice}
\sur{Weiler}}\email{mweiler@mit.edu}

\author[3]{\fnm{Shuhui}
\sur{Qu}}\email{shuhui.qu@samsung.com}

\author[3]{\fnm{Cheol Woo}
\sur{Park}}\email{cheolwoo.p@samsung.com}

\author[2]{\fnm{Tommi}
\sur{S. Jaakkola}}\email{jaakkola@mit.edu}

\author[1]{\fnm{Elsa}
\sur{Olivetti}}\email{elsao@mit.edu}

\author*[1]{\fnm{Rafael}
\sur{Gómez-Bombarelli}}\email{rafagb@mit.edu}

\affil[1]{\orgdiv{Department of Materials Science and Engineering}, \orgname{Massachusetts Institute of Technology}, \orgaddress{\city{Cambridge}, \postcode{02139}, \state{MA}, \country{USA}}}

\affil[2]{\orgdiv{Electrical Engineering \& Computer Science Department}, \orgname{Massachusetts Institute of Technology}, \orgaddress{\city{Cambridge}, \postcode{02139}, \state{MA}, \country{USA}}}

\affil[3]{\orgdiv{Samsung Display America Lab}, \orgaddress{\city{San Jose}, \postcode{95134}, \state{CA}, \country{USA}}}

%%==================================%%
%% Sample for unstructured abstract %%
%%==================================%%

\abstract{
Organic molecular crystals underpin technologies ranging from pharmaceuticals to organic electronics, yet predicting solid-state packing of molecules remains challenging because candidate generation is combinatorial and stability is only resolved after costly energy evaluations. Here we introduce \texttt{PackFlow}, a flow matching framework for molecular crystal structure prediction (CSP) that generates heavy-atom crystal proposals by jointly sampling Cartesian coordinates and unit-cell lattice parameters given a molecular graph. This lattice-aware generation interfaces directly with downstream relaxation and lattice-energy ranking, positioning \texttt{PackFlow} as a scalable proposal engine within standard CSP pipelines. To explicitly steer generation toward physically favourable regions, we propose physics alignment, a reinforcement learning post-training stage that uses machine-learned interatomic potential energies and forces as stability proxies. Physics alignment improves physical validity without altering inference-time sampling. We validate \texttt{PackFlow}'s performance against heuristic baselines through two distinct evaluations. First, on a broad unseen set of molecular systems, we demonstrate superior candidate generation capability, with proposals exhibiting greater structural similarity to experimental polymorphs. Second, we assess the full end-to-end workflow on two unseen CSP blind-test case studies, including relaxation and lattice-energy analysis. In both settings, \texttt{PackFlow} outperforms heuristics-based methods by concentrating probability mass in low-energy basins, yielding candidates that relax into lower-energy minima and offering a practical route to amortize the relax-and-rank bottleneck.

% Across evaluations that include energy relaxations and lattice-energy analysis, \texttt{PackFlow} produces higher-quality proposals than heuristics-based baselines and better concentrates probability mass in low-energy basins. We demonstrate \texttt{PackFlow}'s candidate generation capability on several structural metrics computed on an unseen test set, and the entire end-to-end \texttt{PackFlow}-enabled workflow on 2 CSP blind-test case studies: \texttt{PackFlow} candidates relax into more competitive minima and yield polymorph predictions closer to experiment, offering a practical route to amortizing the relax-and-rank bottleneck in molecular CSP.
}


%%\pacs[JEL Classification]{D8, H51}

%%\pacs[MSC Classification]{35A01, 65L10, 65L12, 65L20, 65L70}

\maketitle

\section{Introduction}\label{sec1}

Organic molecular crystals underpin much of modern molecular materials science, from pharmaceuticals to organic semiconductors \citep{price2008crystal,beran2023frontiers}.
The performance of these materials is determined not solely by the molecular structure, but also by the packing arrangement in the solid state and the specific polymorph that forms under practical processing conditions \citep{lee2011crystal,bernstein2020polymorphism}.
Because polymorphs can exhibit meaningfully different solubility, stability \cite{price2008crystal}, and other functional properties, predicting and controlling solid forms remains a practical necessity across discovery and manufacturing.
Molecular crystal structure prediction (CSP) seeks to infer experimentally realizable crystal packings from minimal chemical information, ideally from a two-dimensional molecular representation.
Over the last two decades, the field has used community-wide, prospective evaluations, most prominently the CSP Blind Tests organized by the Cambridge Crystallographic Data Centre (CCDC) \citep{bardwell2011towards,reilly2016report,hunnisett2024seventh}, to measure progress under controlled conditions.
These benchmarks have helped standardize an end-to-end view of CSP: propose candidate packings across plausible space groups and unit cells, then relax and rank them to identify the most stable structures and near-degenerate competitors that may correspond to observed polymorphs.

Historically, this pipeline has been challenged by two coupled bottlenecks.
First, the search space is combinatorial, spanning lattice parameters, molecular orientations, conformations (for flexible molecules), and space-group symmetry \citep{dudek2022along}.
Second, accurate ranking requires quantum-level energetics, since low-energy structures often differ by only a few kJ mol$^{-1}$ \citep{cruz2015facts,nyman2015static}.
As a result, many workflows rely on hierarchical screening: cheaper models for broad exploration, followed by higher-accuracy refinement, traditionally dispersion-corrected density functional theory (DFT) for final ranking \citep{reilly2016report,zhou2025robust,hoja2018first}.
More recently, machine-learned interatomic potentials (MLIPs) have begun to close the accuracy--cost gap \citep{wood2025family,kovacs2025mace,anstine2025aimnet2}, enabling larger relax-and-rank campaigns while retaining accurate DFT-level energetics that are essential for crystal stability assessment \citep{glick2025toward,gharakhanyan2025fastcsp}.
Within this landscape, methods such as Genarris \citep{tom2020genarris,yang2025genarris} have provided robust proposal engines by generating large, physically filtered pools of randomized crystal candidates to seed downstream optimization \citep{oganov2011evolutionary,case2016convergence,galanakis2024rapid}.
However, these seed-and-relax pipelines continue to encounter scalability challenges.
Generating a proposal set sufficiently diverse to capture relevant low-energy basins often requires extensive enumeration, and reliable ranking still necessitates repeated geometry optimizations under periodic constraints \cite{gharakhanyan2025fastcsp}.

Generative structure prediction via neural networks has emerged as a broad paradigm for translating compact representations into three-dimensional structures, including proteins, molecules, and crystalline materials. Molecular CSP can be viewed as a particularly demanding instance of this general problem.
The landmark success of AlphaFold \citep{jumper2021highly} illustrates what becomes possible when powerful auxiliary signals are available: large sequence databases encode coevolutionary constraints that sharply restrict the space of feasible folds, even without explicitly modeling physical interactions.
In contrast, crystal structure prediction must rely directly on physical interactions under periodic boundary conditions.
Another line of relevant works in generative models for inorganic crystals has demonstrated that learning-based priors can capture features of periodic structure distributions \citep{jiao2023crystal,miller2024flowmm,zeni2025generative}.
However, extending these ideas to molecular crystals is substantially harder because packing is governed by weaker non-covalent interactions and conformational flexibility, which greatly enlarges the accessible configuration space.
Notably, OXtal \citep{jin2025oxtal}, a recent generative molecular CSP approach, demonstrates the promise of directly learning conditional distributions over molecular packing from molecular graphs.
Still, it does not explicitly determine lattice parameters, which precludes direct structure relaxation and energy-based ranking when low-energy candidates must be reliably ordered.

Here we introduce \texttt{PackFlow}, a generative framework for targeting the central computational barrier in scalable molecular CSP: generating high-quality, physically plausible proposals that reduce the downstream relaxation burden without sacrificing coverage of relevant low-energy regions.
First, we perform heavy-atom molecular crystal prediction as a conditional generative task that jointly predicts Cartesian coordinates for heavy atoms in the unit cell, and lattice parameters, enabling immediate use of periodic energy relaxations for ranking.
This joint coordinate–lattice approach both supports post-relaxation energy evaluation needed to resolve near-degenerate polymorphs, and aligns the generator with the periodic boundary conditions used for ranking.
Next, we propose a post-training physics alignment (PA) stage that uses MLIP-derived energies and forces as feedback to steer the proposal distribution toward physically favorable regions of configuration space, thereby amortizing part of the optimization cost that would otherwise be incurred repeatedly during relax-and-rank pipelines.
By combining joint lattice-aware generation with PA post-training, \texttt{\texttt{PackFlow}} is designed as a proposal engine for end-to-end CSP pipelines that is scalable and faithful to the physical criteria that ultimately determine crystal stability. We validate this approach against robust heuristic baselines such as Genarris, through evaluations on a broad unseen test set and two CSP blind-test case studies.
Our results demonstrate that \texttt{PackFlow} generates proposal candidates that not only have better structural fidelity than heuristic methods across a broad set of unseen test crystals, but that also ultimately relax into lower-energy minima as demonstrated on two unseen CSP blind-test case studies. To this end, \texttt{PackFlow} offers a practical route to amortize the relax-and-rank bottleneck.

\section{Results}\label{sec2}

\subsection{\texttt{\texttt{PackFlow}}-enabled crystal structure prediction workflow}
\begin{figure}[ht!]
\centering
\makebox[\linewidth][c]{%
  \includegraphics[width=1.25\linewidth]{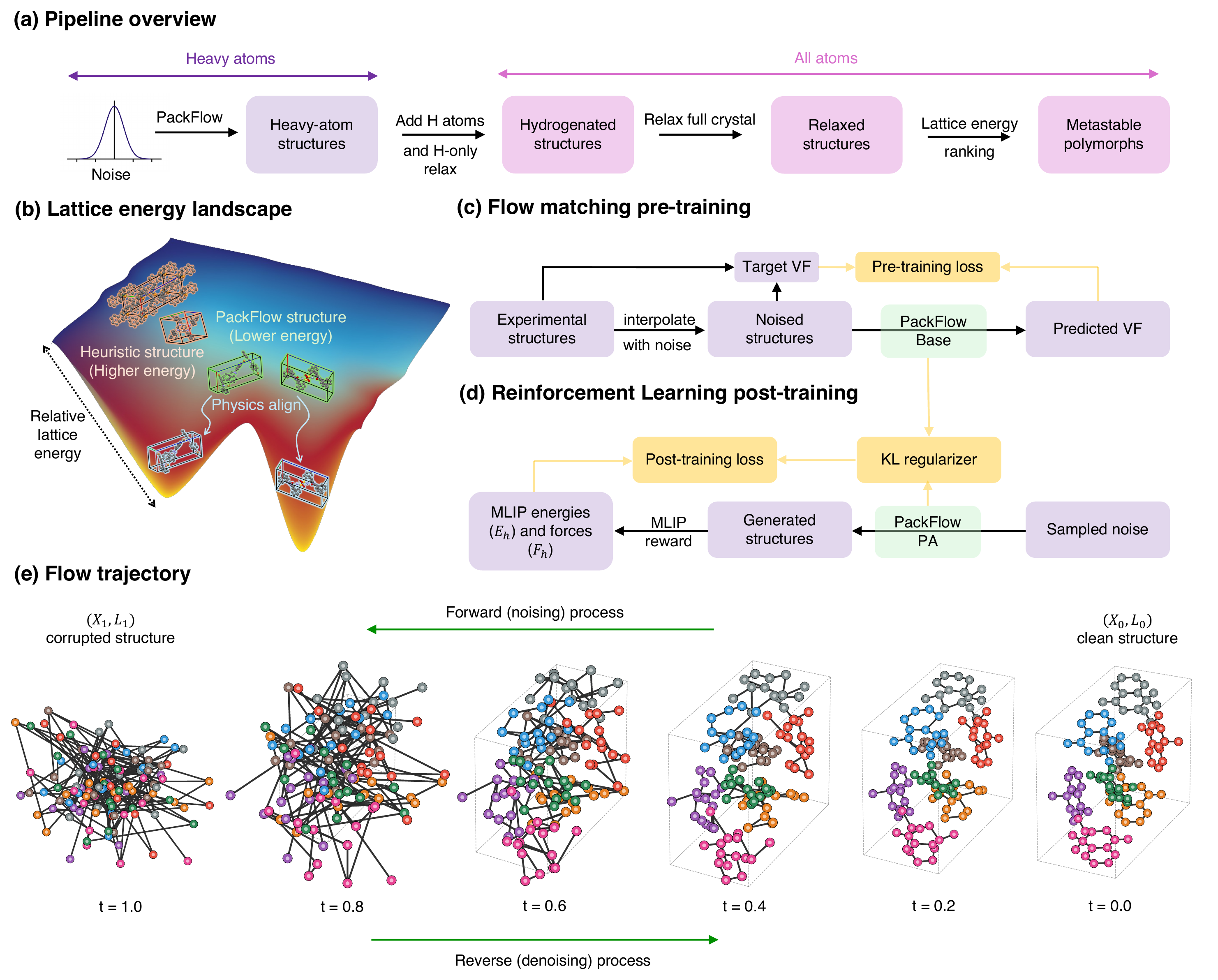}%
}
\caption{\textbf{Schematic overview of \texttt{\texttt{PackFlow}}-enhanced molecular crystal structure prediction workflow.} \textbf{(a)} Pipeline consists of several steps: heavy atom crystal generation with \texttt{PackFlow}, Hydrogen addition and H-only relaxation with MLIP, and full crystal relaxation and lattice energy ranking with MLIP to obtain final metastable polymorphs. \textbf{(b)} \texttt{PackFlow-Base} models generate candidates with lower lattice energies than heuristic structure generation methods. Physics alignment drives lattice energies further down in comparison to base model initialization. \textbf{(c, d)} \texttt{PackFlow} training is divided into pre-training and post-training stages. Pre-training (c) involves simultaneous training of coordinate and lattice with standard flow-matching objective. Post-training (d) involves training with GRPO objective on heavy-atom energies and forces (approximate) obtained from MLIP. We abbreviate ``vector-field" as VF. \textbf{(e)} A sample flow trajectory on the test set demonstrating joint sampling of lattice and coordinates. }
\label{fig:Figure1}
\end{figure}

Fig.~\ref{fig:Figure1} summarizes the end-to-end workflow that we evaluate in this work. We frame \emph{heavy-atom} molecular crystal prediction as a generative task: given a molecular graph (atom types and covalent bonds), we generate (i) Cartesian atomic coordinates for all heavy atoms in the unit cell and (ii) lattice parameters. These heavy-atom proposals are then completed into full crystal structures by adding hydrogens and performing a short hydrogen-only relaxation, followed by full crystal relaxation and lattice-energy ranking with a machine-learned interatomic potential (MLIP) (Fig.~\ref{fig:Figure1}a). This staged protocol reflects a practical CSP setting where generative models are used to propose candidates that are subsequently relaxed and ranked by an energy model. We restrict the generative modeling component to the candidate generation stage, rather than targeting an end-to-end replacement of the entire workflow since these models currently do not guarantee robustness in out-of-distribution (OOD) settings \cite{subramanian2024closing}, and hence will often require energetic refinements.

\texttt{PackFlow} is trained in two stages: flow-matching pre-training, followed by physics-alignment post-training (Fig.~\ref{fig:Figure1}c,d). On unseen examples, the pre-trained \texttt{\texttt{PackFlow}} produces heavy-atom candidates whose downstream relaxed lattice energies are already improved relative to random or heuristics-based structure generation procedures (Fig.~\ref{fig:Figure1}b). We further introduce a reinforcement learning post-training stage (physics alignment) that uses MLIP-derived energetic and force-based feedback to perform small local atomic and lattice adjustments and steer the pre-trained generator further toward physically favorable regions of configuration space (Fig.~\ref{fig:Figure1}b). Physics alignment shifts the distribution of generated candidates toward lower lattice-energy structures compared to the pre-trained initialization (Fig.~\ref{fig:Figure1}b). In the remainder of this section, we first establish how the \texttt{\texttt{PackFlow}} architecture and training choices affect geometric fidelity and packing statistics (Figs.~\ref{fig:Figure2}--\ref{fig:Figure3}, Table~\ref{tab:Table1}), and then quantify how physics alignment improves structural quality (Fig.~\ref{fig:Figure4}, Table~\ref{tab:Table2}), and finally demonstrate improved CSP outcomes using all \texttt{PackFlow} models (Table~\ref{tab:Table2}, Fig.~\ref{fig:Figure5}). To distinguish between training stages, we denote the pre-trained model as \texttt{PackFlow-Base} and the post-trained version as \texttt{PackFlow-PA} (Fig.~\ref{fig:Figure1}c).

% \begin{figure}[h!]
% \centering
% % \includegraphics[width=1\textwidth]{figures/Figure1.pdf}
% \includegraphics[width=1.2\textwidth]{figures/Figure1.pdf}

\subsection{A bond-aware flow-matching transformer}
\label{sec:flow_matching}
\subsubsection{Problem setup and flow-matching objective}
A molecular crystal with $N$ heavy atoms is represented by atom types $\{a_i\}_{i=1}^N$, Cartesian coordinates $x \in \mathbb{R}^{N\times 3}$, and lattice parameters $\ell \in \mathbb{R}^6$ (lengths and angles). Angles are represented using the unconstrained representation following \citet{miller2024flowmm}. \texttt{PackFlow} learns a time-dependent vector field conditioned on atom types and covalent bonding graph, that maps noisy states $(x_t,\ell_t)$ to a vector-field that transports noise to data via an ordinary differential equation (ODE). Using the optimal-transport (OT) interpolation schedule, we construct noisy states by
\begin{align}
x_t &= (1-t_x)\,x_0 + t_x\,\varepsilon_x, \qquad \varepsilon_x \sim \mathcal{N}(0, I), \\
\ell_t &= (1-t_\ell)\,\ell_0 + t_\ell\,\varepsilon_\ell, \qquad \varepsilon_\ell \sim \mathcal{N}(0, I),
\end{align}
where $(x_0,\ell_0)$ denotes a training crystal and $t_x,t_\ell \in [0,1]$ are flow times for coordinates and lattice, respectively. Note that in our notation, we use $t = 0$ for clean data, and $t = 1$ for noise, following our implementation. Under this schedule, the corresponding target vector fields are
\begin{align}
v_x^\star(x_t,t_x) &= \varepsilon_x - x_0, \\
v_\ell^\star(\ell_t,t_\ell) &= \varepsilon_\ell - \ell_0,
\end{align}
which follow from $\frac{d}{dt}\big((1-t)u + t z\big) = z-u$. \texttt{PackFlow} minimizes a flow-matching \cite{lipman2022flow} regression loss
\begin{align} \label{eq:flow_matching_loss}
\mathcal{L}_{\text{FM}}(\theta)
&=
\mathbb{E}\!\left[
\frac{1}{N}\left\lVert v_{x,\theta}(x_t,\ell_t,t_x,t_\ell)-v_x^\star\right\rVert_2^2
+
\lambda_{\ell}\left\lVert v_{\ell,\theta}(x_t,\ell_t,t_x,t_\ell)-v_\ell^\star\right\rVert_2^2
\right],
\end{align}
where $v_{x,\theta}$ and $v_{\ell,\theta}$ are neural predictions, and $\lambda_{\ell}$ is a lattice-loss weight. Sampling is performed by integrating the learned ODEs backward in time from Gaussian initial conditions ($t=1$) to $t=0$, producing heavy-atom coordinates and lattice parameters that are subsequently post-processed as in Fig.~\ref{fig:Figure1}a.

\subsubsection{Evaluation of heavy-atom proposal quality}
\label{sec:metric_overview}

Although the flow-matching objective in Eq.~\eqref{eq:flow_matching_loss} trains \texttt{PackFlow} to match the joint coordinate--lattice data distribution, assessing a molecular CSP proposal engine requires complementary metrics that probe different failure modes. We therefore report metrics covering (i) global unit-cell plausibility, (ii) local physical validity, (iii) proximity to the experimental polymorph, (iv) distributional packing agreement, and (v) computational cost. We briefly summarize each below; formal definitions appear in Section~\ref{sec:eval_metrics}.

\textbf{Global plausibility: Density error.}
\emph{Density error} measures how well generated unit-cell volumes (with the correct composition) reproduce experimental crystal densities. Because density is a compact proxy for packing tightness, it reveals systematic \emph{under-} or \emph{over-compression} prior to relaxation. Lower density error indicates more realistic global packing.

\textbf{Local validity: Clash rate.}
\emph{Clash rate} quantifies severe short-range heavy-atom overlaps under periodic boundary conditions using a covalent-radius-based threshold. It serves as a fast proxy for whether a proposal is physically realistic, and is a viable initialization for hydrogen addition and MLIP relaxation: low clash rates indicate respect for steric constraints, whereas high clash rates imply large forces and unstable downstream optimization.

\textbf{Target proximity: AMD distance.}
To measure closeness to the ground-truth polymorph (up to periodicity), we use the \emph{average-minimum-distance (AMD)} descriptor distance \cite{widdowson2020average, widdowson2022resolving}. AMD provides a robust periodic representation of the induced point pattern; lower AMD indicates closer agreement while remaining origin-invariant and stable to small perturbations. We use AMD as our primary closeness-to-target metric for heavy-atom proposals.

\textbf{Packing agreement: RDF Wasserstein distances.}
Beyond target matching, we evaluate \emph{distributional} packing statistics via Wasserstein distances between radial distribution function (RDF) histograms of heavy-atom proposals and experimental ground truths. The full-range RDF Wasserstein distance captures agreement in global intermolecular distance patterns, while the \emph{short-range} variant focuses on local contact geometry where steric repulsion and covalent/hydrogen-bonding distances typically manifest. Together, these metrics test whether generated structures reproduce realistic distance distributions.

\textbf{Cost: Wall-clock generation time.}
Finally, we report the wall-clock time per heavy-atom proposal (ODE solve only) to quantify inference throughput under a fixed sampling budget. This timing excludes downstream post-processing (hydrogenation and relaxations), isolating the cost of candidate generation itself.

\noindent
Taken together, these metrics distinguish proposals that are (i) physically valid yet far from the experimental polymorph, (ii) globally plausible but locally pathological, and (iii) both realistic and close to the target at low generation cost. Full implementation details are provided in Section~\ref{sec:eval_metrics}.

% \begin{figure}[ht!]
% \centering
% \includegraphics[width=1\textwidth]{figures/Figure2.pdf}
\begin{figure}[ht!]
\centering
\makebox[\linewidth][c]{%
  \includegraphics[width=1.25\linewidth]{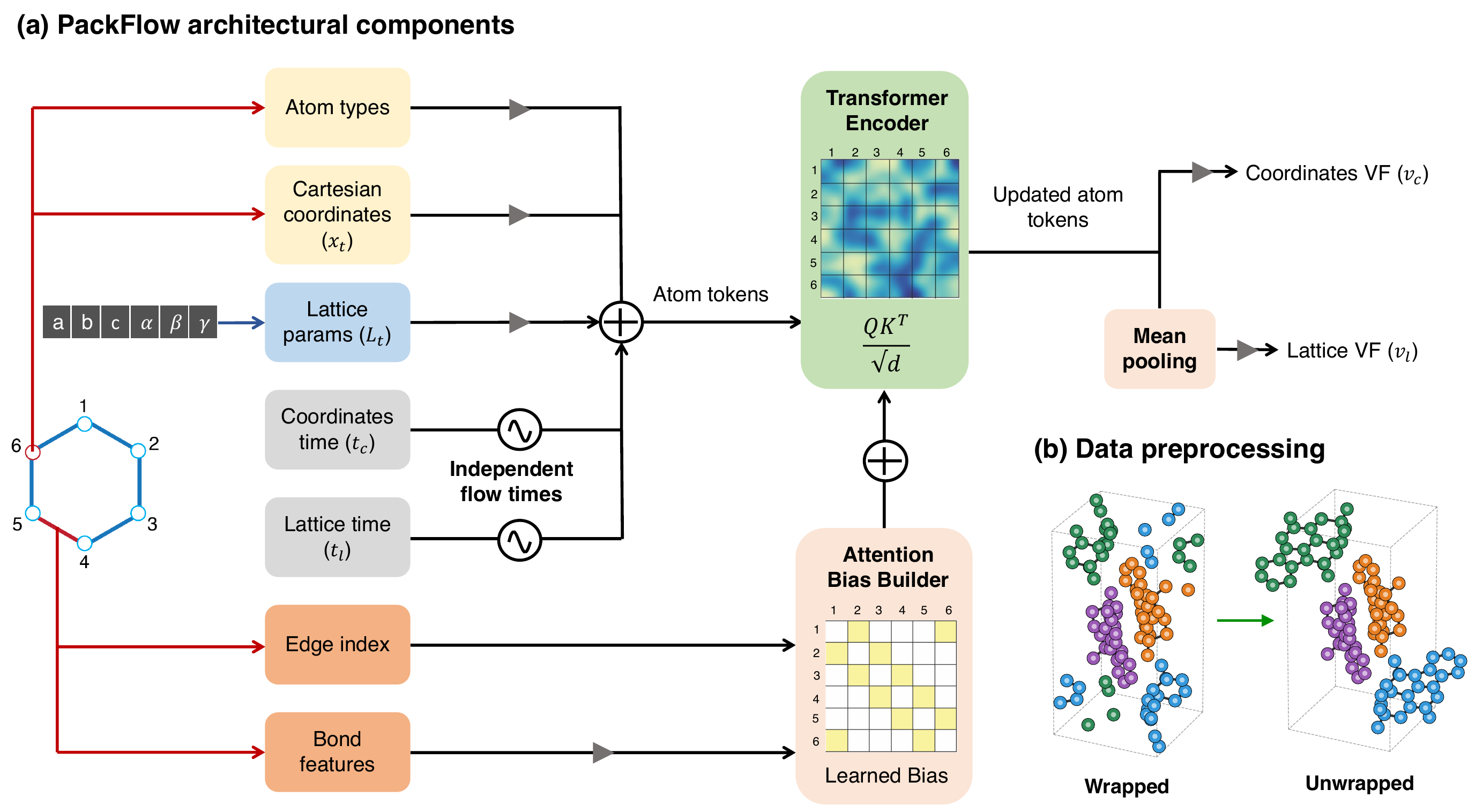}%
}
\caption{\textbf{Architectural components of \texttt{PackFlow}, and crystal preprocessing.} \textbf{(a)} Coordinates, atom types, lattice parameters, and flow times are jointly embedded into per-atom tokens, which are fed into transformer encoder. Coordinate and lattice flow times are sampled independently. Covalent bonding information is embedded as an additive attention bias to the transformer attention scores. Coordinate and lattice vector fields (VFs) are obtained as readouts from updated atom tokens. \textbf{(b)} For all crystals in the dataset, molecules are made whole at unit cell boundaries (unwrapped) and centered by mean subtraction. See Section~\ref{sec:unwrapping} for details.}
\label{fig:Figure2}
\end{figure}

\subsubsection{Independent flow times for coordinates and lattice}
\texttt{PackFlow} embeds coordinates, lattice parameters, and flow times into per-atom tokens and processes them with a transformer encoder (Fig.~\ref{fig:Figure2}a). A key design choice is to sample \emph{independent} flow times $t_x$ and $t_\ell$ for the coordinate and lattice channels, rather than forcing a shared $t$. Intuitively, unit-cell degrees of freedom can require a different denoising schedule than local atomic rearrangements. Independent time parameterization allows the model to learn distinct progression dynamics for packing and lattice geometry.

This choice yields measurable gains. When we ablate independent times by training a model with shared coordinate/lattice times, performance degrades across most structural metrics (Table~\ref{tab:Table1}, “Shared”): density error increases and structural similarity metrics worsen relative to the reference \texttt{PackFlow-Base} model that uses independent times. These results indicate that decoupling coordinate and lattice times is beneficial even when the architecture and training budget are held fixed. While this idea of independent flow times has to the best of our knowledge, not been applied to lattices and coordinates, it has recently found success in joint flow of backbone $C_{\alpha}$ coordinates, and per-residue latent variables for protein generation \cite{geffner2025proteina}.    

\subsubsection{Covalent-bond attention bias improves physical validity}
A second key choice is how the model incorporates molecular connectivity. \texttt{PackFlow} injects covalent bonding information as a Graphormer-style \cite{ying2021transformers} \emph{additive attention bias} to the transformer’s attention logits (Fig.~\ref{fig:Figure2}a). Concretely, for attention head $h$, the pre-softmax attention score between atoms $i$ and $j$ is modified as
\begin{align}
s_{ij}^{(h)} = \frac{\langle q_i^{(h)}, k_j^{(h)}\rangle}{\sqrt{d_h}} + \beta^{(h)}(b_{ij}),
\end{align}
where $q_i^{(h)}$ and $k_j^{(h)}$ are query/key projections, $d_h$ is the head dimension, and $\beta^{(h)}(\cdot)$ is a learned function of bond features $b_{ij}$ (with a learnable head-wise baseline). This bias encourages the model to attend preferentially along covalent edges without restricting attention to only bonded pairs, preserving the ability to model long-range packing interactions. More details on attention bias construction are provided in Section~\ref{sec:attention_bias}.

The ablation study in Table~\ref{tab:Table1} shows that explicit bond information is critical for physical validity. Removing bonding (“No Bonds”) signal completely from the input representation increases the heavy-atom clash rate dramatically (from 2.74\% to 19.01\%) and degrades geometric similarity metrics (AMD and RDF distances). This is expected since lack of covalent bonding information renders molecular CSP non-unique since any isomer containing the specified atoms would be a valid generation choice. This demonstrates that simple re-purposing of inorganic CSP methods will result in poor performance. Replacing attention bias with a bond-GNN message passing module (“Bonds-GNN”) partially recovers performance, but does not match the overall trade-off achieved by attention bias under the same training budget. 

This is a plausible outcome because the GNN injects bonding information only indirectly through node embeddings before the transformer, whereas an additive attention bias provides an explicit \emph{pairwise} bond-dependent term in the attention logits at \emph{every} transformer layer. The latter gives the model a more direct and repeatedly available mechanism to prioritize covalently bonded neighbors during attention while still permitting global (non-bonded) interactions. Together, these results support attention bias as an effective and parameter-efficient mechanism to encode covalent constraints while still allowing global packing interactions. More architectural details can be found in Section~\ref{sec:arch_details}.

% This again is an expected outcome, since performing a GNN forward pass and integrating bonding signal as per-token priors is less flexible than having an attention-biasing mechanism that is directly integrated into each transformer layer. 

\subsubsection{Unwrapped data representation} \label{sec:unwrapping}
As shown in Fig.~\ref{fig:Figure2}b, as a data-preprocessing step, we \emph{unwrap} each molecule across periodic boundaries using covalent connectivity before converting to Cartesian coordinates.
This avoids discontinuities that arise when bonded atoms are wrapped to opposite faces of the unit cell (i.e., bond vectors differing by integer lattice translations), yielding smoother bond geometry and pairwise-distance statistics for the model to learn. This unwrapping is performed so that a side for unwrapping is chosen canonically to be the side that contains the molecular centroid, allowing the majority of each molecule to always lie inside the unit cell. Using an unwrapped representation was found to be an essential design choice, since our initial attempts with fractional coordinates and wrapped cartesian coordinates (similar to recent inorganic CSP methods such as \cite{miller2024flowmm}) were unable to train effectively. More details on the exact unwrapping algorithm used and other symmetry considerations are provided in Sections~\ref{sec:unwrapping} and \ref{sec:symmetries}.

\begin{table}[htbp!]
\centering
\caption{\textbf{Ablation studies on bond information and time embedding for heavy-atom crystal generation.}
The reference model (\texttt{PackFlow}-Base) is shown at the top which uses attention bias as the method for bond inclusion, and independent flow times for lattice and coordinate flows. The ``Bonds-GNN" and ``No Bonds" rows refer to \texttt{PackFlow}-Base models with covalent bonding information included using a GNN embedding layer, and no bonding information, respectively. ``Shared" row refers to model trained with shared sampling of  flow times for both coordinate and lattice flows. All models shown in this table were trained using the 60M size model on a single NVIDIA V100 GPU for 4 days. Note that these are shorter ablation runs and therefore have different metrics than 60M entry in Table~\ref{tab:Table2}, which was trained on 4 V100 GPUs for 4 days. Metric definitions are provided in Section~\ref{sec:eval_metrics}. Best values across all models are highlighted in \textbf{bold}.}
\label{tab:Table1}

\small
\setlength{\tabcolsep}{4pt}
\renewcommand{\arraystretch}{1.15}

\begin{tabular}{lccccc}
\toprule
\textbf{Model} &
\textbf{Density} $\downarrow$ &
\textbf{Clash} $\downarrow$ &
\textbf{AMD} $\downarrow$ &
\textbf{RDF} $\downarrow$ &
\textbf{RDF Wass} $\downarrow$ \\
& \textbf{Error \%} & \textbf{\%} & \textbf{$L_\infty$} & \textbf{Wass} & \textbf{(Short)} \\
\midrule

\textbf{\texttt{PackFlow-Base}} & \textbf{5.20} & \textbf{2.74} & \textbf{0.293} & \textbf{0.067} & 0.118 \\
\midrule

\multicolumn{6}{l}{\textit{Bond Inclusion}} \\
Bonds-GNN & 5.83 & 3.65 & 0.320 & 0.067 & \textbf{0.117} \\
No Bonds  & 5.88 & 19.01 & 0.332 & 0.070 & 0.134 \\
\midrule

\multicolumn{6}{l}{\textit{Flow Time}} \\
Shared & 6.17 & 3.54 & 0.312 & 0.068 & 0.119 \\
\bottomrule
\end{tabular}
\end{table}

\subsection{Matching experimental statistics and symmetric attention}
\begin{figure}[ht!]
\centering
\makebox[\linewidth][c]{%
  \includegraphics[width=1.25\linewidth]{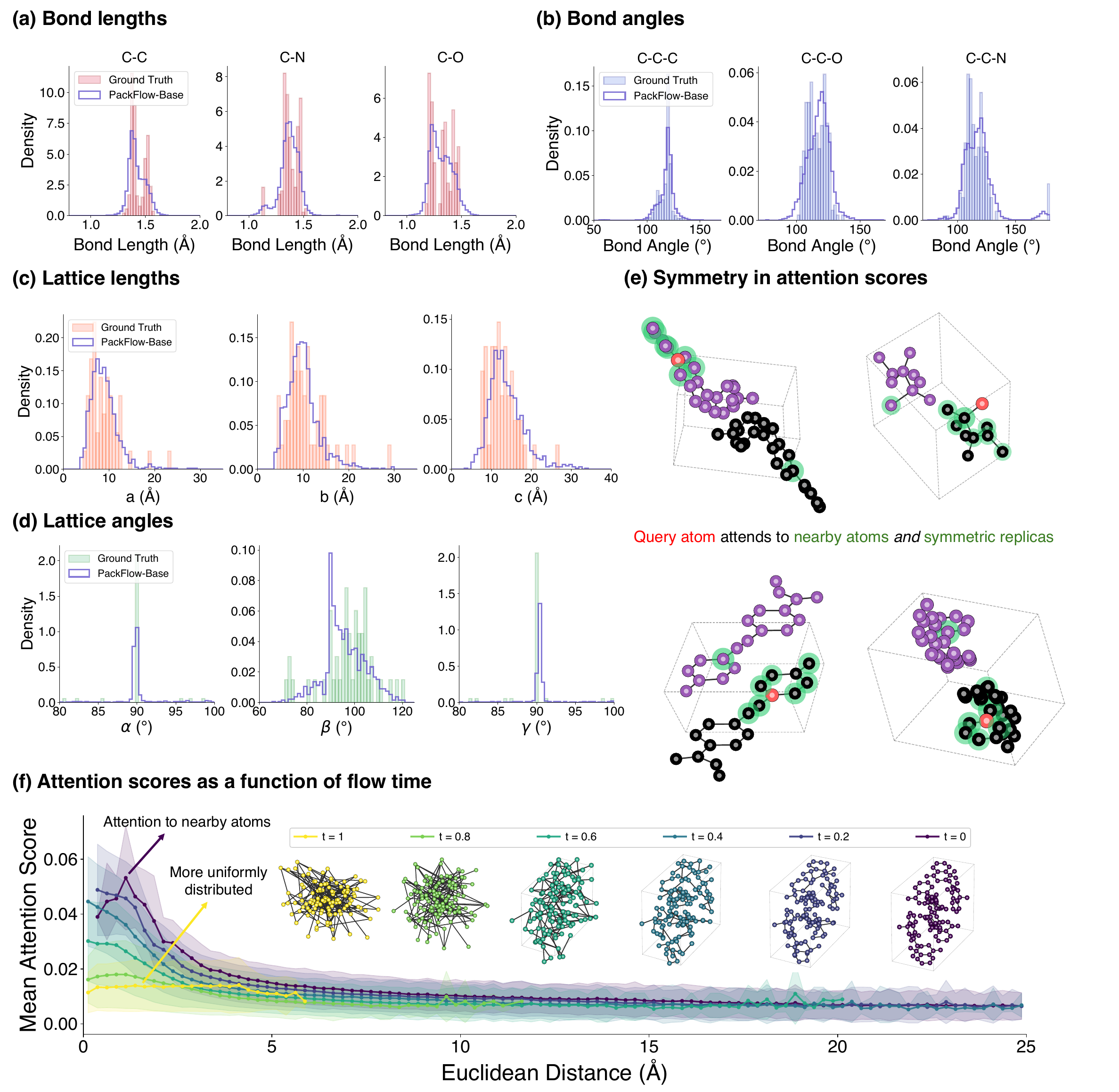}%
}
\caption{\textbf{Generation quality of \texttt{PackFlow}-Base models, and analysis of learned attention scores.} \textbf{(a)} Bond lengths, \textbf{(b)} bond angles, \textbf{(c)} lattice lengths, \textbf{(d)} lattice angles distributions of generated crystals on test data in comparison to experimental ground truth distributions. \textbf{(e)} Atoms learn to attend to nearby neighbors and (possibly) distant atoms in different molecules that are symmetric replicas. ``Query" atoms are colored red, and ``key" atoms are colored green. \textbf{(f)} Average pairwise attention score profile is uniformly spread out at high flow times, and peaks at small distances at low flow times, indicating transition from global to local attention with flow progression. }
\label{fig:Figure3}
\end{figure}

We next evaluate whether the pre-trained generator reproduces key geometric statistics of experimentally determined crystals. \texttt{PackFlow-Base} closely matches ground-truth distributions of local molecular geometry, including bond lengths and bond angles (Fig.~\ref{fig:Figure3}a,b), as well as lattice lengths and angles (Fig.~\ref{fig:Figure3}c,d). These agreement trends are consistent with the low structural discrepancy errors in Table~\ref{tab:Table1} and Table~\ref{tab:Table2} (AMD and RDF Wasserstein distances, and density error), indicating that the model captures both intramolecular geometry and intermolecular packing statistics. All metric definitions are provided in Section~\ref{sec:eval_metrics}.

To better understand how the model coordinates local chemistry with global packing, we analyze learned attention patterns (Fig.~\ref{fig:Figure3}e,f). At the atom-pair level, attention maps emphasize both near neighbors, \emph{and} pairs that are far apart in Cartesian space yet systematically related by the crystal symmetry within the same unit cell. For instance, Fig.~\ref{fig:Figure3}e shows that the query atom (red) not only attends to nearest intramolecular atoms, but also to a symmetrically equivalent \textit{intermolecular} atom. In molecular crystals, such symmetry-related replicas often arise because one molecular site generates an orbit under the space-group operations (i.e., the symmetry mates of a Wyckoff site), producing multiple copies of the molecule within the unit cell. While \texttt{PackFlow} does not explicitly enforce space-group symmetry (unlike, for example, \cite{kazeev2025wyckoff, puny2025space} in inorganic crystals), the presence of consistently high attention between these related regions suggests that the model can approximately capture recurring symmetry-induced packing relationships present in the training data. 

Aggregating attention as a function of pair distance reveals a pronounced flow time dependence (Fig.~\ref{fig:Figure3}f): at high flow times (early in generation), attention is relatively delocalized, whereas at low flow times (late in generation) attention concentrates at short distances. This transition is consistent with a generation process that first organizes coarse global packing context and then progressively refines local atomic arrangements to avoid short-range clashes. Lastly, we observe a peak that uniquely appears as $t \rightarrow 0$ flow time at short Euclidean distances, showing that there is a tendency for the model to attend strongly at certain equilibrium separation distances, and taper off at more repulsive and attractive regimes.

% \begin{figure}[t]
% \centering
% \includegraphics[width=1\textwidth]{figures/Figure3.pdf}

\subsection{Physics alignment via reinforcement learning}

\subsubsection{Alignment signal from heavy-atom energies and forces}
Pre-training with the flow-matching objective in Eq.~\eqref{eq:flow_matching_loss} emphasizes distribution matching and geometric fidelity, but does not guarantee crystal stability or clash-free generations. We therefore introduce physics alignment (PA) post-training for \texttt{PackFlow} (Fig.~\ref{fig:Figure4}a). This approach is formulated as an RL problem:  An \textit{agent} (\texttt{PackFlow}) takes \textit{actions} (sampling), and receives rewards via interaction with the \textit{environment} (MLIP). PA post-training uses an MLIP to score each generated heavy-atom proposal by (i) heavy-atom energy $E_h$ and (ii) a scalar heavy-atom force statistic $F_h$ (implemented as a per-structure force-norm mean). These quantities provide computationally tractable proxies for all-atom stability signals prior to hydrogen addition and full relaxation (see Section~\ref{sec:proxy_rewards} for more details on this approximation). We then use these scores to post-train a policy with a group relative policy optimization (GRPO) objective \cite{shao2024deepseekmath}.

% Starting from a pre-trained reference policy (\texttt{PackFlow}-Base with parameters $\theta_{\mathrm{ref}}$), use rewards derived from $E_h$ and/or $F_h$ to post-train the model with a group relative policy-optimization (GRPO) objective \cite{shao2024deepseekmath}. 

The following sections highlight key algorithmic design choices (in comparison to previous works applying RL to diffusion models \cite{chen2025accelerating, park2025guiding} and stochastic interpolants \cite{hoellmer2026open}) to ensure RL post-training is tractable for our flow matching-based generator. Extended discussions of the methodological details can be found in Section~\ref{sec:rl_methods}.

\subsubsection{Group relative policy optimization for flow models}
% To physics-align \texttt{PackFlow}, we post-train the pre-trained generator with group relative policy optimization (GRPO) \cite{shao2024deepseekmath}. GRPO is well-suited to crystal structure prediction because it compares multiple candidate packings \emph{for the same conditioning context} (same molecule and unit-cell contents) and learns from \emph{relative} quality within that group. This avoids relying on absolute reward scales that are not comparable across different crystals (e.g., extensive energies and molecule-dependent force scales).
We post-train the pre-trained generator \texttt{PackFlow-Base} with group relative policy optimization (GRPO) \cite{shao2024deepseekmath}. In our setting, we explicitly construct each GRPO ``group'' to contain multiple candidate packings generated under the \emph{same conditioning context} (i.e., the same molecule and unit-cell contents). With this grouping choice, GRPO learns from \emph{within-context} comparisons: it encourages candidates that score better than other candidates for the same crystal specification, rather than requiring a globally calibrated notion of what constitutes a ``good'' reward across different molecules or unit cells. This is convenient for crystal structure prediction because absolute reward magnitudes are often not directly comparable across crystals (e.g., due to extensive energies and molecule-dependent force scales), whereas relative preferences within a fixed context are more meaningful and stable.

\textbf{Setup and rewards.}
For each crystal template, we sample a group of $K$ heavy-atom packings from the current policy and score each sample using MLIP-based proxy signals (heavy-atom energy and a force statistic; Fig.~\ref{fig:Figure4}a). We convert these proxy scores into within-group standardized advantages (and mix multiple reward channels at the advantage level rather than mixing raw rewards). The policy update increases the likelihood of higher-advantage samples while penalizing drift from the pre-trained reference via a KL regularizer, yielding stable post-training that preserves geometric fidelity.

\textbf{Adapting GRPO to flow matching models.}
A key difference from prior GRPO applications is that \texttt{PackFlow} is an ODE-based flow matching model rather than an autoregressive policy or a diffusion model. Standard proximal policy optimization (PPO)-style updates require a tractable per-sample likelihood (or an equivalent scoring quantity) to form importance ratios and KL penalties. For continuous-time normalizing flows, endpoint likelihoods can be computed via an integral of the vector-field divergence along the sampling trajectory \cite{chen2018neural,grathwohl2018ffjord}, but evaluating this exactly during RL post-training is computationally expensive.

To make GRPO practical at scale for flow matching, we introduce a \emph{single-time surrogate} that approximates the needed per-sample score using only one sampled time (and the associated flow-matching terms), rather than integrating quantities along the full trajectory. This preserves the \emph{relative} learning signal required by GRPO (since group advantages depend on within-group comparisons) while reducing the RL inner-loop cost substantially. We detail the surrogate construction, the resulting importance-ratio form, and the KL estimator in Section~\ref{sec:rl_methods}.

\textbf{ODE sampling choice.}
We choose ODE sampling primarily for efficiency and simplicity. In practice, deterministic ODE samplers can reach comparable sample quality with fewer function evaluations than SDE-based samplers, making them faster in wall-clock time \cite{song2020score,song2020denoising}. This matters in our RL setting because rollouts are generated at every outer-loop iteration (Algorithm~\ref{alg:grpo_packflow}), so sampling speed directly impacts training cost. ODE sampling also reduces manual design choices during rollouts: unlike SDE-based samplers, it does not require specifying an explicit stochastic noise scale and schedule (e.g., a $\sigma(t)$ schedule) or injecting noise throughout the trajectory \cite{hoellmer2026open}.

\subsubsection{Advantage mixing instead of reward mixing}
\label{sec:adv_mixing}
A practical challenge in multi-objective post-training is that energy- and force-derived rewards ($r_E$ and $r_F$) can differ substantially in scale and variability. Naively mixing raw rewards (e.g., $r = \lambda r_E + (1-\lambda) r_F$, where $\lambda$ is a scaling coefficient) typically requires hand-tuning scaling factors so that neither objective dominates updates \cite{moskovitz2023confronting}.

We instead mix normalized, group relative advantages rather than raw rewards. Concretely, using per-sample rewards
$r_E^{(k)}=-E_h^{(k)}$ and $r_F^{(k)}=-F_h^{(k)}$, we first standardize each objective within the same group:
\begin{align}
\tilde A_E^{(k)} &= \frac{r_E^{(k)}-\mu_E}{\sigma_E+\epsilon_{\mathrm{adv}}},\qquad
\tilde A_F^{(k)} = \frac{r_F^{(k)}-\mu_F}{\sigma_F+\epsilon_{\mathrm{adv}}},
\end{align}
where $\mu$ and $\sigma$ denote within-group mean and standard deviation, and $\epsilon_{\mathrm{adv}}$ is a numerical stabilizer. We then mix these normalized advantages and apply clipping:
\begin{equation}
\label{eq:grpo_adv}
A_{\lambda}^{(k)}
=
\mathrm{clip}\!\left(
\lambda\,\tilde A_E^{(k)} + (1-\lambda)\,\tilde A_F^{(k)},
\,-c,\,c
\right),
\qquad \lambda\in[0,1].
\end{equation}
Because $\tilde A_E$ and $\tilde A_F$ are standardized, Eq.~\eqref{eq:grpo_adv} provides a \emph{scale-free} interpolation between energy- and force-dominant alignment regimes without any manual reward (Fig.~\ref{fig:Figure4}c). This reduces sensitivity to MLIP-dependent reward magnitudes, while still allowing $\lambda$ to smoothly control the energy--force trade-off (Fig.~\ref{fig:Figure4}d). The pseudocode can be found in Algorithm \ref{alg:grpo_adv}.

\subsubsection{Alignment improves physical validity and proximity to ground truths}
\begin{figure}[ht!]
\centering
\makebox[\linewidth][c]{%
  \includegraphics[width=0.97\linewidth]{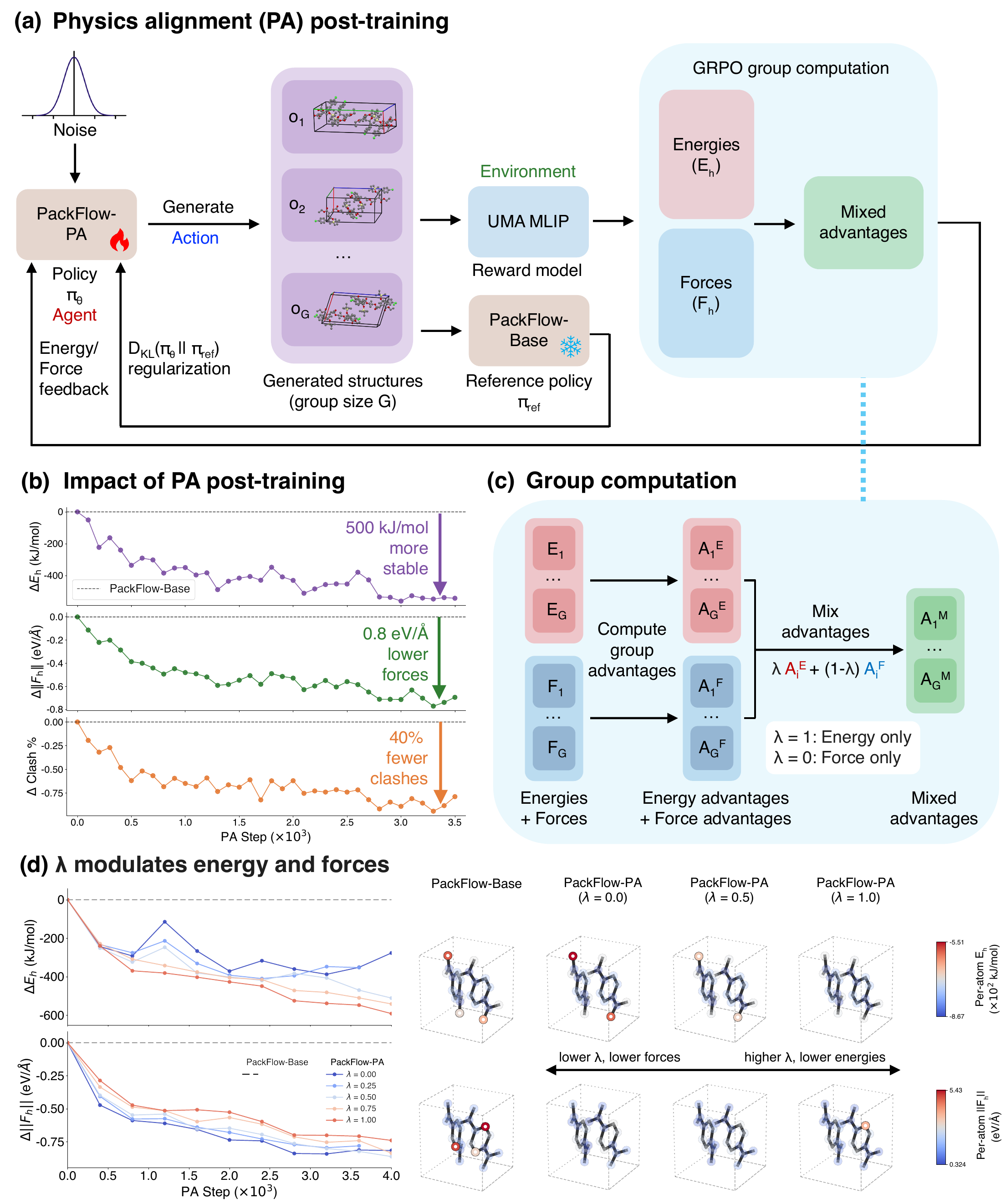}%
}
    \caption{\textbf{Physics alignment (PA) post-training approach. } \textbf{(a)} \texttt{PackFlow}-PA is trained by generating a set of observations $O_i$ from the current policy, evaluating heavy-atom energies $E_h$ and forces $F_h$ (approximations to all-atom energies and forces) using MLIP, mixing energy and force advantages using advantage mixing (details in (c)), and using the advantages as feedback to update the policy. KL divergence regularization term is computed between current policy and reference policy (\texttt{PackFlow}-Base) to ensure that post-training does not steer far away from base model distribution. \textbf{(b)} post-training results in significant lowering of $E_h$ and $F_h$, and reduction in atomic clashes on test data. \textbf{(c)} Advantage mixing linearly interpolates advantages instead of rewards for multi-objective post-training. Advantages being normalized quantities, do not require manual re-scaling which is typically required if rewards of varying scales are mixed directly. \textbf{(d)} Variation of mixing parameter $\lambda$ allows smooth tradeoff in test performance between $E_h$ and $F_h$. Curves of test performance as a function of PA step are shown on left, and an example structure colored by per-atom energies and forces as a function of $\lambda$ on the right. Red-/blue- colored atoms indicate higher/lower $E_h$/$F_h$, respectively. }
\label{fig:Figure4}
\end{figure}

Physics alignment (PA) yields improvements in heavy-atom energy and force-based proxy objectives on the unseen test set, over the course of post-training (Fig.~\ref{fig:Figure4}b), and simultaneously reduces the frequency of unphysical atomic clashes (Fig.~\ref{fig:Figure4}b).

The gains from alignment translate to improved structure metrics on the test set at fixed inference cost. Table~\ref{tab:Table2} compares \texttt{PackFlow} model scales and physics-aligned variants against Genarris baselines. Starting from \texttt{PackFlow-60M}, physics alignment reduces clash rates (from 2.53\% to 1.53--1.71\% depending on $\lambda$) and yields consistent improvements in distributional similarity metrics (RDF distances) and/or structural proximity (AMD), with \emph{no} additional sampling cost because the architecture and ODE solve are unchanged (Table~\ref{tab:Table2}, identical wall times for \texttt{PackFlow-60M} and \texttt{PackFlow-PA}).

Because PA is defined by a multi-objective MLIP stability proxy (energy and a force-based strain signal), varying the mixing weight $\lambda$ simply shifts which aspect of the same proxy is emphasized during post-training. Varying $\lambda$ produces a smooth trade-off between prioritizing lower energies and prioritizing lower force (Fig.~\ref{fig:Figure4}d, left). Visually, we observe an interpolation of per-atom $E_h$ and $F_h$ as $\lambda$ evolves (Fig.~\ref{fig:Figure4}d, right), indicating that energy- and force-focused alignment are compatible but not identical objectives in practice. 

We also see this trade-off reflected in structural metrics. As shown in Table~\ref{tab:Table2}, emphasizing forces more strongly (smaller $\lambda$) yields the lowest clash rates (best at $\lambda=0$), while emphasizing energies (larger $\lambda$) improves structural proximity to experimental ground truth as measured by AMD (best at $\lambda=1$). The hybrid setting $\lambda=0.5$ provides a robust compromise, remaining near-optimal across clash rate, AMD, and RDF-based distances. Overall, these results support physics alignment as a practical post-training approach that improves physical plausibility and structural match to experimental ground truths beyond the base model. In practice, we find that the pre-trained model is responsible for most of the global structure formation, producing coherent, physically reasonable packings, while post-training predominantly applies targeted local refinements (e.g., correcting short-range contacts and subtle lattice/coordinate perturbations) to mitigate remaining unphysical artifacts.

% We also see this interpolation playing a role on downstream structural metrics, i.e. it can be seen from Table~\ref{tab:Table2} that interestingly, $\lambda = 0.5$ manages the best tradeoff between metrics such as Clash \% and AMD. These results support physics alignment as a practical post-training approach that improves physical plausibility and downstream stability beyond the base model.

% We observe neither forces-only ($\lambda=0$) nor energy-only ($\lambda=1$) perform best across the metrics. 
% Interestingly, the energy-force hybrid setting ($\lambda=0.5$) yields the best holistic performance across structural fidelity (Table~\ref{tab:Table2}), and energetics (Table~\ref{tab:TableG1}). Specifically, it achieves the best tradeoff between structural metrics: AMD, RDF Wasserstein, density error, clash, and energetics-based metrics, i.e., energies and forces. This indicates the synergistic effect of leveraging \textit{multiple} rewards for PA post-training i.e., \textit{both} energy and forces. 

% \begin{figure}[h!]
% \centering
% \includegraphics[width=0.91\textwidth]{figures/Figure4.pdf}

\begin{table}[htbp!]
\centering
\caption{\textbf{Effect of model scale and physics-alignment on heavy-atom crystal generation.}
We compare three base model sizes and the PA models with various mixing parameters ($\lambda$) against Genarris baselines. PA models were trained using \texttt{PackFlow-60M} as their initialization. PA models exhibit lower clash rates, and higher structural similarity to ground truth structures than the base models, while maintaining the same inference wall-clock time. Base and PA models also outperform Genarris models on most structural metrics, and inference speed. Best values are \textbf{bolded} and second-best values are \underline{underlined}.}
\label{tab:Table2}

\small
\setlength{\tabcolsep}{4pt}
\renewcommand{\arraystretch}{1.15}

\begin{tabular}{lcccccc}
\toprule
\textbf{Model} &
\textbf{Density} $\downarrow$ &
\textbf{Clash} $\downarrow$ &
\textbf{AMD} $\downarrow$ &
\textbf{RDF} $\downarrow$ &
\textbf{RDF Wass} $\downarrow$ &
\textbf{Wall} $\downarrow$ \\
& \textbf{Error \%} & \textbf{\%} & \textbf{$L_\infty$} & \textbf{Wass} & \textbf{(Short)} & \textbf{Time (s)} \\
\midrule

\multicolumn{7}{l}{\textit{Baselines}} \\
Genarris Plain & 27.89 & \textbf{0.28} & 0.903 & 0.114 & 0.211 & 0.120 \\
Genarris Rigid Press & 19.77 & 1.93 & 0.541 & 0.071 & 0.118 & 0.312 \\
\midrule

\multicolumn{7}{l}{\textit{Ours (Base)}} \\
\texttt{PackFlow-Base-2M}  & 5.89 & 9.38 & 0.367 & 0.074 & 0.149 & \textbf{0.063} \\
\texttt{PackFlow-Base-20M} & 5.30 & 2.66 & 0.291 & 0.066 & 0.114 & \underline{0.087} \\
\texttt{PackFlow-Base-60M} & \textbf{4.69} & 2.53 & 0.286 & 0.065 & 0.114 & 0.128 \\
\midrule

\multicolumn{7}{l}{\textit{Ours (Physics-Aligned)}} \\
\texttt{PackFlow-PA} ($\lambda=0$)   & 5.91 & \underline{1.53} & 0.284 & 0.063 & 0.105 & 0.128 \\
\texttt{PackFlow-PA} ($\lambda=0.5$) & 5.39 & 1.62 & \underline{0.278} & \textbf{0.063} & \textbf{0.104} & 0.128 \\
\texttt{PackFlow-PA} ($\lambda=1$)   & \underline{5.21} & 1.71 & \textbf{0.277} & \underline{0.063} & \underline{0.104} & 0.128 \\
\bottomrule
\end{tabular}
\end{table}

\subsection{Blind-test case studies show proximity to experimental polymorphs}
% Finally, we examine the end-to-end CSP behavior on two unseen CSP blind-test examples (Fig.~\ref{fig:Figure5}), \texttt{OBEQOD} from the 5th CSP Blind test \cite{bardwell2011towards} being a rigid test case (tier 1 in \cite{zhou2025robust}), while \texttt{XAFPAY01} from the 6th CSP Blind test \cite{reilly2016report} being a more challenging flexible test case (tier 3 in \cite{zhou2025robust}). 

\begin{figure}[ht!]
\centering
\makebox[\linewidth][c]{%
  \includegraphics[width=1.25\linewidth]{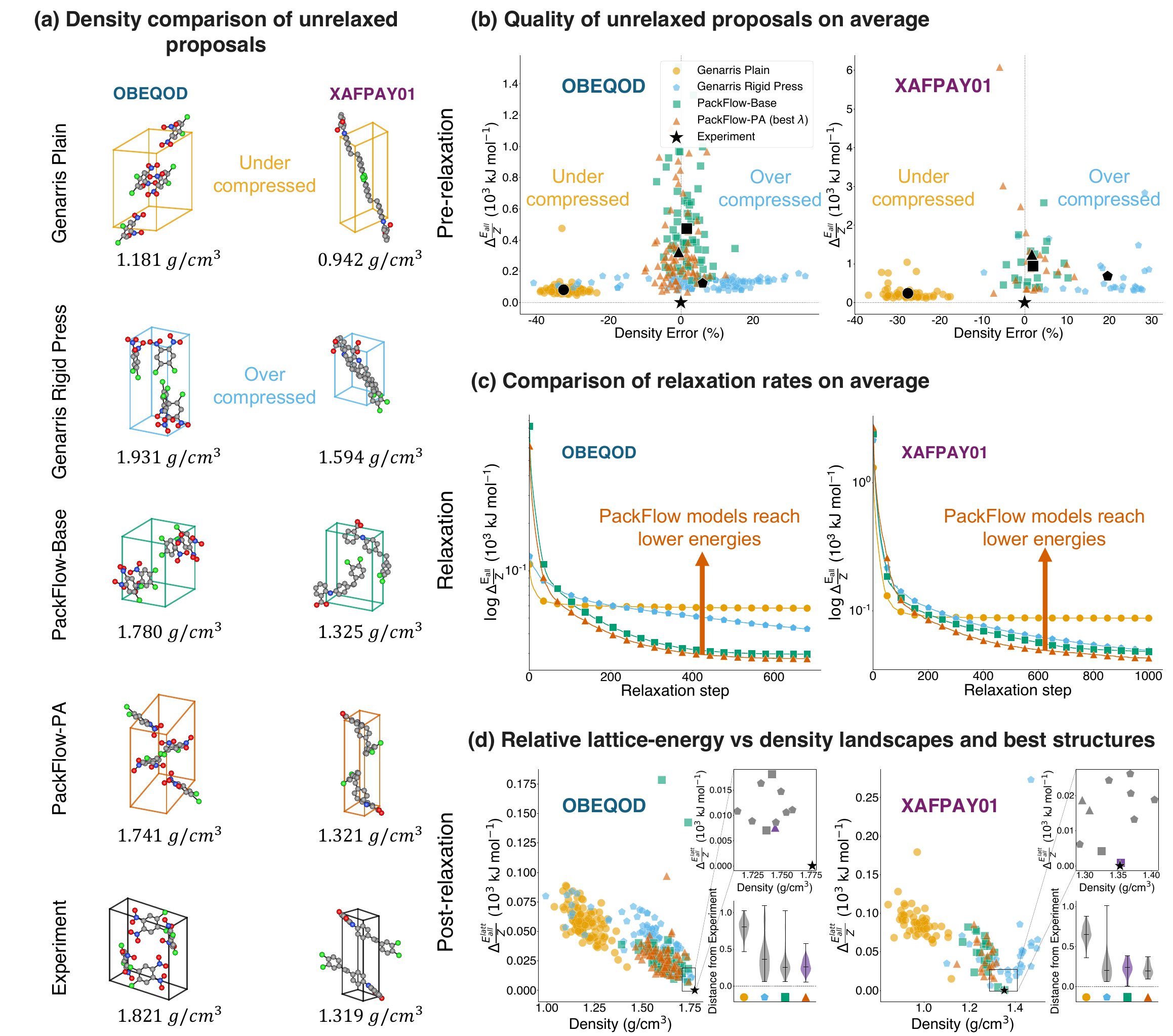}%
}
\caption{\textbf{Comprehensive comparison of \texttt{PackFlow} and Genarris on two CSP blind test examples.} \textbf{(a)} \texttt{PackFlow} models generate initial proposals that are closer in density to experimental structure, than Genarris baselines. Genarris Plain typically under-compresses, while Genarris Rigid Press over-compresses structures relative to experiment. Structures with median density among predicted samples are visualized. \textbf{(b)} Both Genarris methods tend to produce proposals with lower energies on average than \texttt{PackFlow} models prior to relaxation, but with under-compression and over-compression in densities with respect to experiment. \textbf{(c)} \texttt{PackFlow} structures reach lower lying minima than Genarris structures after MLIP relaxation. \textbf{(d)} After relaxation, \texttt{PackFlow} models result in polymorphs that are closer to experiment in density/lattice-energy space, than Genarris polymorphs. Purple colors in zoomed and violin subplots indicate minimum distance in relative-lattice-energy-density plane to experimental polymorph; the ``best" predictions across methods are used for comparison here, rather than averages. All relative energies across figures are of hydrogenated crystals, and are calculated with respect to the fully relaxed structure with the global minimum energy. Points in all figures were after application of a bond-length filter to remove implausible molecular geometries. More details in Section~\ref{sec:fig5_vis_filters}.}
\label{fig:Figure5}
\end{figure}

Finally, we examine end-to-end CSP behavior on two unseen CSP blind-test examples (Fig.~\ref{fig:Figure5}). Five crystals across the CSP blind tests satisfied our dataset preprocessing criteria (see Section~\ref{sec:CCDC_preprocess}); we therefore focus on two representative cases spanning rigidity and flexibility: \texttt{OBEQOD} from the 5th CSP Blind Test \cite{bardwell2011towards} (rigid; tier~1 in \cite{zhou2025robust}) and \texttt{XAFPAY01} from the 6th CSP Blind Test \cite{reilly2016report} (flexible; tier~3 in \cite{zhou2025robust}). 

Prior to any energy relaxation, \texttt{PackFlow} proposals are substantially closer in density to the experimental structures than Genarris baselines (Fig.~\ref{fig:Figure5}a). Consistent with the qualitative depiction in Fig.~\ref{fig:Figure5}a, Genarris Plain tends to under-compress while Genarris Rigid Press tends to over-compress, whereas \texttt{PackFlow} samples concentrate nearer to experimental densities. Genarris methods can yield lower pre-relaxation energies on average (Fig.~\ref{fig:Figure5}b), but their density biases and associated structural distortions lead to less favorable relaxed minima. 

Energy trends can change upon relaxation \cite{day2005third}. During MLIP relaxation, \texttt{PackFlow}-generated candidates reach lower-lying minima than Genarris-generated candidates (Fig.~\ref{fig:Figure5}c). \texttt{PackFlow-PA} reaches an even lower energy minima compared to the base model due to physics alignment post-training. 
Notably, within just 100 samples, the best \texttt{PackFlow} polymorphs are closer to experiment in the density--relative-lattice-energy plane (Fig.~\ref{fig:Figure5}d) than Genarris polymorphs, with relaxed lattice-energy differences to the experimental polymorph on the order of only a few~kJ/mol (one case slightly above the canonical $\sim$5~kJ/mol CSP target, and the other well within it; Fig.~\ref{fig:Figure5}d).

% Notably, within just 100 samples, the best \texttt{PackFlow} polymorphs are closer to experiment in the density--relative-lattice-energy plane (Fig.~\ref{fig:Figure5}d) than Genarris polymorphs. 

These case studies illustrate that improving the \emph{proposal distribution}, in particular by producing physically plausible packings with realistic densities and fewer short-range conflicts, can yield better relaxed outcomes even when baseline methods occasionally start with lower unrelaxed energies.

% Overall, across quantitative test-set metrics (Tables~\ref{tab:Table1}--\ref{tab:Table2}) and blind-test examples (Fig.~\ref{fig:Figure5}), \texttt{PackFlow} combines (i) bond-aware generation that respects intramolecular chemistry, (ii) flexible coordinate/lattice denoising via independent time parameterizations, and (iii) MLIP-guided physics alignment via reinforcement learning to produce crystal candidates that are both structurally faithful and energetically favorable for downstream CSP pipelines.

% \begin{figure}[ht!]
% \centering
% \makebox[\linewidth][c]{%
%   \includegraphics[width=\linewidth]{figures/Figure5.pdf}%
% }

\FloatBarrier
\section{Discussion}\label{sec:discussion}

\texttt{PackFlow} is a generative framework for molecular crystal structure prediction that targets the \emph{proposal} stage of end-to-end CSP by \emph{jointly} sampling heavy-atom coordinates and unit-cell lattice parameters. This joint generation is a practical advantage over recent coordinate-only approaches such as OXtal \cite{jin2025oxtal}: by producing a complete periodic structure, \texttt{PackFlow} can interface directly with periodicity-sensitive MLIPs/DFT for either/both post-training and downstream relax-and-rank, which is essential to identify stable polymorphs from proposals. Moreover, purely generative proposals often lack fidelity in out-of-distribution settings \cite{subramanian2024closing}; in these cases, energetic refinement through relaxations is especially valuable. 
% In this work, we focus comparisons on baselines with publicly available, end-to-end reproducible implementations at the time of writing; this excludes OXtal from runnable benchmarking.
% We did not include OXtal as a runnable baseline in our experiments because, at the time of writing, the OXtal preprint does not point to a publicly available implementation/checkpoints or data splits needed to reproduce training/inference.

% Across the full unseen test set, \texttt{PackFlow-Base} produces proposals with substantially more realistic unit-cell densities than heuristic baselines, corresponding to up to an 83\% reduction in density error relative to Genarris baselines, while maintaining fast sampling ($\sim$0.1\,s per structure on a single V100 GPU). These proposal-level gains translate into improved relaxed outcomes: in the blind-test case studies, \texttt{PackFlow} candidates more consistently relax into lower-energy minima and yield structures closer to experiment in the density--relative-lattice-energy plane (Fig.~\ref{fig:Figure5}). 
Across the full unseen test set, \texttt{PackFlow-Base} produces proposals with substantially more realistic unit-cell densities than heuristic baselines, corresponding to up to an 83\% reduction in density error relative to Genarris baselines, while maintaining fast sampling ($\sim$0.1\,s per structure on a single V100 GPU). These proposal-level gains translate into improved relaxed outcomes: in the blind-test case studies, \texttt{PackFlow} candidates more consistently relax into lower-energy minima and yield structures closer to experiment in the density--relative-lattice-energy plane (Fig.~\ref{fig:Figure5}), with relaxed lattice energies within only a few~kJ/mol of the experimental polymorph in both cases (near the commonly cited $\sim$5~kJ/mol CSP accuracy target).

We further show that post-training with physics alignment can improve physical validity without changing inference-time sampling. Starting from \texttt{PackFlow-Base}, physics alignment yields up to an additional 40\% reduction in clash rate relative to \texttt{PackFlow-Base} at identical wall-clock generation time (Table~\ref{tab:Table2}), and the aligned models better concentrate probability mass in physically favorable regions, which carries through relaxation in the blind-test analyses. 

Two modeling choices were particularly important for making joint coordinate--lattice generation reliable in this heavy-atom setting: encoding covalent connectivity with attention bias (critical for avoiding unphysical structures) and decoupling coordinate and lattice denoising dynamics via independent time parameterization (Table~\ref{tab:Table1}).
Overall, \texttt{PackFlow} is already ready to deploy in practical CSP workflows as a drop-in replacement for heuristic proposal engines, offering substantial improvements in proposal quality under the same relax-and-rank pipeline.

There are several limitations and extensions remaining for future work. First, we restricted our dataset to \emph{homomolecular} crystals, similar to some CSP works such as \citet{gharakhanyan2025fastcsp}, and \citet{galanakis2024rapid}. While the modeling components introduced here are not specific to single-component systems, extending \texttt{PackFlow} to co-crystals and solvates will require additional treatment to ensure that learned packing statistics remain well calibrated across chemically heterogeneous pairs and solvent inclusion. 

% A second, orthogonal limitation is that our flow-matching formulation conditions on the number of molecules in the unit cell ($Z$) being known \emph{a priori}. In this work, we compare \texttt{PackFlow} and Genarris baselines on equal footing by providing the true $Z$ for each target crystal. However, in practical CSP settings $Z$ is often unknown and must be inferred or explored. An important direction for future work is to augment generative CSP pipelines with mechanisms to \emph{predict or sample} $Z$ jointly with coordinates and lattice parameters, enabling fully end-to-end proposal generation without assuming unit-cell content is given.

Second, to make GRPO-based post-training tractable, we approximated alignment rewards using heavy-atom proxies (heavy-atom energies and force norms) rather than full all-atom signals after hydrogen addition and hydrogen-only relaxation. We show that these proxies are strongly correlated with downstream stability (see Section~\ref{sec:proxy_rewards}), however they inevitably diminish reward fidelity in cases where hydrogens strongly influence short-range contacts, electrostatics, or subtle packing preferences. We chose this approximation because performing hydrogen completion and relaxation at every GRPO step is substantially more expensive and would increase both wall-clock time and memory pressure during training, especially when maintaining multiple samples per template for group relative updates. A natural extension is to use \emph{multi-fidelity} alignment schemes that combine frequent proxy rewards with occasional full all-atom evaluations. 
In addition, while dense self-attention has $O(N^2)$ compute in the number of tokens $N$ and is standard in modern transformers, its \emph{memory} footprint need not be quadratic with optimized kernels such as \texttt{FlashAttention} \cite{dao2022flashattention}, which avoid materializing the full $N\times N$ score matrix. In our implementation, the dominant scaling bottleneck is instead the \emph{explicit} head-wise attention-bias tensor, which is constructed as a dense $[B,n_{\text{head}},N,N]$ mask and therefore materializes $O(N^2)$ VRAM and incurs substantial memory I/O (low arithmetic intensity), limiting the benefits of fused attention kernels. Consequently, increasing the atom count per unit cell can reduce feasible batch sizes and slow both pre-training and post-training. Future work could mitigate this by redesigning bond conditioning to avoid dense bias materialization (e.g., sparse/block-sparse or low-rank bias parameterizations compatible with \texttt{FlashAttention}-style implementations).

% In addition, the current transformer backbone has an inherent scaling limitation: dense self-attention (and, in our implementation, an explicit head-wise attention-bias tensor) scales roughly quadratically in the maximum number of tokens per crystal. Consequently, increasing the atom count per unit cell increases attention compute and memory footprint superlinearly, which can reduce feasible batch sizes and slow training/post-training. Future work could for instance mitigate this bottleneck using sparse or block-sparse attention.

% Fourth, this work is intended as a proof-of-concept that demonstrates that improved proposals can translate into better downstream relaxed outcomes on held-out crystals, including CSP blind-test examples. Our evaluation uses a limited sampling budget (100 proposals per crystal), whereas large-scale CSP studies often require substantially more candidates to estimate hit rates and robustly characterize polymorph coverage. A comprehensive structural match analysis (for example, a systematic evaluation against experimental targets under standardized matching criteria) would require generating and relaxing many more structures per crystal, which is computationally intensive and beyond the scope of the present study. We therefore focused on metrics that have been used to assess CSP proposal quality, including AMD distance to experimental polymorph, density accuracy, clash rates, distributional similarity, and proximity in density--relative-lattice-energy space, as a practical proxy for end-to-end usefulness under a constrained relaxation budget.

Third, our end-to-end CSP evaluation follows a standard relax-and-rank protocol in which generated proposals are locally minimized and then ordered by relaxed MLIP lattice energies. This implicitly assumes that low-energy basins under the chosen energy model are a useful proxy for experimentally realized polymorphs, even though crystallization outcomes can depend on kinetics, solvent-mediated pathways, and finite-temperature free-energy effects. At the same time, because \texttt{PackFlow} is trained on experimentally reported CSD/CCDC structures, the learned proposal distribution reflects a form of \emph{survivorship bias} toward packings that were experimentally accessible and characterizable, which can partially encode kinetic accessibility in the training signal. Coupling this learned prior with energy-based relaxation tends to be most faithful for relatively rigid molecules, where conformational degrees of freedom are limited and lattice-energy rankings more often correlate with observed forms; for flexible molecules, additional uncertainty arises because the crystallized conformer and pathway-dependent trapping can shift which polymorphs appear, so our rankings should be interpreted as thermodynamics-leaning within the experimentally biased proposal set rather than a complete kinetic model.

Finally, our models were trained at moderate scale (up to 60M parameters) and on a limited dataset ($\sim$80k crystals) to establish feasibility. The scaling trends observed with model size suggest that additional compute and data could further improve geometric fidelity, reduce clash rates, and increase the fraction of candidates that relax into low-energy minima. Scaling \texttt{PackFlow} along these axes, together with richer training distributions (for example, broader chemical coverage and more diverse unit-cell contents), represents a clear direction for improving generalization and increasing the practical impact of generative CSP in large-scale polymorph discovery workflows.

\section{Methods}

\subsection{Architecture}
\label{sec:arch_details}

Fig.~\ref{fig:Figure2}a summarizes the high-level dataflow; here we document the architectural components as reflected by our implementation.

\subsubsection{Per-atom tokenization}
\texttt{PackFlow} represents each heavy atom $i$ by a width-$d$ token formed by summing a set of modality-specific embeddings:
\begin{equation}
h_i
=
\phi_x(x_{t,i})
\;+\;
\phi_{\text{node}}(f_i)
\;+\;
\phi_\ell(\ell_t)
\;+\;
\phi_{t_x}(t_x)
\;+\;
\phi_{t_\ell}(t_\ell),
\label{eq:token_sum}
\end{equation}
where $\phi_x:\mathbb{R}^3\!\to\!\mathbb{R}^d$ is an MLP applied to Cartesian coordinates, $\phi_{\text{node}}$ is an MLP applied to per-atom descriptors $f_i$ (e.g., atom identity and RDKit-derived features), $\phi_\ell:\mathbb{R}^6\!\to\!\mathbb{R}^d$ embeds the (time-dependent) lattice state, and $\phi_{t_x},\phi_{t_\ell}$ embed the coordinate and lattice flow times via sinusoidal features followed by MLPs.
Importantly, $\phi_\ell(\ell_t)$ and the time embeddings are broadcast to all atoms in the same crystal (indexed by the batch map), so the transformer can condition each local update on global unit-cell context at every layer.
Equation~\eqref{eq:token_sum} is purely a representation detail: training still minimizes the flow-matching objective in Eq.~\eqref{eq:flow_matching_loss}.

\subsubsection{Encoding covalent connectivity} \label{sec:attention_bias}
Covalent bonds are incorporated in one of two mutually exclusive ways.

\textbf{Additive attention bias.} We build a head-specific bias tensor from per-edge bond features and add it to the pre-softmax attention logits in \emph{every} transformer layer (as described in Section~\ref{sec:flow_matching} and depicted in Fig.~\ref{fig:Figure2}a). Operationally, this is implemented by (1) mapping each bond feature vector to $n_{\text{head}}$ scalars, (2) scattering these scalars into a dense $[B,n_{\text{head}},M,M]$ bias (with a learnable per-head baseline), and (3) passing the resulting additive mask into multihead self-attention. This preserves global attention while systematically increasing attention mass along covalent edges.

\textbf{Bond-GNN.} As an ablation-compatible  (see Table~\ref{tab:Table1}), we apply several GATv2Conv \cite{brody2021attentive} layers on the sparse covalent graph \emph{before} dense batching, using the same bond features as edge attributes, and then feed the updated node states into the transformer. This injects bonding information via message passing rather than attention reweighting; the rest of the architecture is unchanged.

\subsubsection{Readouts for coordinate and lattice vector fields}
After transformer processing, \texttt{PackFlow} predicts the coordinate flow with a per-token MLP head:
\[
\hat v_x(x_t,\ell_t,t_x,t_\ell)_i = \psi_x(h_i^{\text{out}})\in\mathbb{R}^3.
\]
For the lattice flow, we mean-pool atom tokens within each crystal and apply a lattice head:
\[
\hat v_\ell(x_t,\ell_t,t_x,t_\ell) = \psi_\ell\!\Big(\mathrm{meanpool}\{h_i^{\text{out}}\}_{i\in\text{crystal}}\Big)\in\mathbb{R}^6.
\]
Both heads are trained under the same objective in Eq.~\eqref{eq:flow_matching_loss}, using the per-atom and per-crystal normalizations defined in our implementation.

\subsection{Physics alignment of flow matching via reinforcement learning}
\label{sec:rl_methods}

We post-train the pre-trained \texttt{PackFlow-Base} model using GRPO with a PPO-style clipped objective. We emphasize the following key ingredients of our approach and direct readers to the original papers for more detailed background: (i) group relative advantages computed from MLIP rewards, (ii) a tractable \emph{single-time surrogate} for a per-sample policy score based on Eq.~\eqref{eq:flow_matching_loss} that is used to construct both importance ratio as well as KL divergence terms of the objective (Equation~\ref{eq:grpo}) and (iii) multi-objective alignment where we mix \emph{normalized} advantages (rather than raw rewards) as defined in Eq.~\eqref{eq:grpo_adv}. We term this post-training approach \emph{physics alignment} (PA), and its pseudocode can be found in Algorithm \ref{alg:grpo_packflow}.

% \paragraph{Relation to prior RL for diffusion and velocity-based generative models.}
% Several recent works apply PPO-style RL to diffusion or stochastic-interpolant samplers by identifying the policy with the reverse-time transition kernel and using explicit Gaussian conditionals to compute per-step likelihood ratios along the full trajectory, typically with KL regularization to a frozen reference model \citep{arxiv2511_03112,arxiv2511_07158}. Another recent line develops inference-time RL for velocity-field models by adding controlled stochasticity to otherwise deterministic dynamics in order to define per-step transition distributions and enable likelihood-ratio estimation \citep{arxiv2602_00424}. In \texttt{PackFlow}, sampling is a deterministic ODE and exact endpoint likelihoods would require divergence integration along the trajectory, which is expensive in an inner-loop RL update. We therefore adopt a single-time flow-matching surrogate that yields a cheap relative policy score for completed samples, allowing GRPO-style post-training while keeping inference-time sampling unchanged.

\subsubsection{Sampling and rewards}
For each crystal template, we sample a group of $K$ heavy-atom proposals
$\{(x_0^{(k)},\ell_0^{(k)})\}_{k=1}^K$
by integrating the learned ODE sampler. 
GRPO operates on groups defined by a single crystal template $\tau$ (fixed molecule/graph and unit-cell contents).
For each $\tau$, we sample $K$ candidates $\{(x_0^{(k)},\ell_0^{(k)})\}_{k=1}^K$ and compute rewards (e.g., $r_E^{(k)}=-E_h^{(k)}$ and/or $r_F^{(k)}=-F_h^{(k)}$).
We then standardize rewards within the same group to obtain group-relative advantages that rank polymorphs of the same crystal.
We do not mix different crystals in a group, since energy/force magnitudes are not comparable across different compositions and sizes (e.g., energy is extensive and polymorph energy scales differ by molecule), which would confound optimization.

Rewards are computed from MLIP evaluations on the generated heavy-atom structures. In our experiments we consider:
\begin{align}
r_E^{(k)} &= -E_h^{(k)},\\
r_F^{(k)} &= -F_h^{(k)},
\end{align}
where $E_h^{(k)}$ is the heavy-atom energy and $F_h^{(k)}$ is a scalar heavy-atom force summary returned by the MLIP (implemented as a per-structure force-norm mean). Sampling is treated as part of the environment interaction: during optimization, $(x_0^{(k)},\ell_0^{(k)})$ are treated as constants and no gradients are propagated through the sampling procedure.

\subsubsection{Single-time surrogate score for importance-ratio and KL}
\label{sec:single_time_surrogate}
Exact endpoint log-likelihoods for flow models would require a divergence integral (or equivalent Hutchinson estimator) along the sampling ODE, which is expensive. Instead, we use a tractable \emph{single-time} surrogate score defined by the integrand of the flow matching objective in Eq.~\eqref{eq:flow_matching_loss} evaluated at a single time/noise draw.

For each sampled proposal $(x_0^{(k)},\ell_0^{(k)})$, we draw a scalar time
$t^{(k)}\sim \mathcal{U}(0,1)$
and Gaussian noises $(\varepsilon_x^{(k)},\varepsilon_\ell^{(k)})\sim\mathcal{N}(0,I)$.
We use OT interpolation to stay consistent with the noise schedule used during pre-training (see Section~\ref{sec:flow_matching}). We form the noisy states and targets
\begin{align}
x_t^{(k)} &= (1-t_x^{(k)})\,x_0^{(k)} + t_x^{(k)}\,\varepsilon_x^{(k)}, \\
\ell_t^{(k)} &= (1-t_l^{(k)})\,\ell_0^{(k)} + t_l^{(k)}\,\varepsilon_\ell^{(k)}, \\
u_x^{(k)} &= \varepsilon_x^{(k)} - x_0^{(k)}, \\
u_\ell^{(k)} &= \varepsilon_\ell^{(k)} - \ell_0^{(k)}.
\end{align}
We then define the per-sample surrogate loss as
% \begin{equation}
% \label{eq:single_time_loss}
% \ell_\theta^{(k)}
% =
% \frac{1}{N}\left\lVert v_{x,\theta}(x_t^{(k)},\ell_t^{(k)},t_x^{(k)},t_l^{(k)})-u_x^{(k)}\right\rVert_2^2
% +
% \lambda_{\ell}\left\lVert v_{\ell,\theta}(x_t^{(k)},\ell_t^{(k)},t_x^{(k)},t_l^{(k)})-u_\ell^{(k)}\right\rVert_2^2.
% \end{equation}
\begin{equation}
\label{eq:single_time_loss}
\ell_\theta^{(k)}(t^{(k)})
=
\frac{1}{N}\left\lVert v_{x,\theta}(x_t^{(k)},\ell_t^{(k)},t_x^{(k)},t_\ell^{(k)})-u_x^{(k)}\right\rVert_2^2
+
\lambda_{\ell}\left\lVert v_{\ell,\theta}(x_t^{(k)},\ell_t^{(k)},t_x^{(k)},t_\ell^{(k)})-u_\ell^{(k)}\right\rVert_2^2.
\end{equation}

We treat $\ell_\theta^{(k)}$ as a practical proxy for $-\log \pi_{\theta}$
up to an unknown affine transformation. This is not an exact endpoint likelihood for the deterministic ODE sampler, but an empirical approximation choice. The pseudocode for this single-time surrogate loss can be found in Algorithm \ref{alg:single_time_loss}. This will later be used in both importance-ratios as well as KL terms, in both cases appearing as a \emph{difference} of two terms. For the sake of simplicity we use \emph{shared} time sampling instead of \emph{independent} time sampling during post-training:
\begin{equation}
t_x^{(k)} = t_\ell^{(k)} \equiv t^{(k)}.
\end{equation}
During optimization, the same frozen $(t^{(k)},\varepsilon_x^{(k)},\varepsilon_\ell^{(k)})$ are reused when evaluating $\ell_\theta^{(k)}$ across PPO epochs for that rollout batch. 

% Concretely, while diffusion-based RL updates that form trajectory log-probabilities by summing per-step transition log-densities, determistic ODEs like used in \texttt{PackFlow} would require expensive per-timestep computation of divergence/traces for likelihoods. To avoid this, we define a single cheap surrogate likelihood for the completed sample under frozen $(t,\varepsilon)$, which we reuse across PPO epochs for stability.

\textbf{Importance-ratio.} For each rollout batch, we cache $\ell_{\theta_{\mathrm{old}}}^{(k)}$ once using the parameters $\theta_{\mathrm{old}}$ that generated the batch (i.e., before the PPO inner loop). Within the inner loop, we take multiple PPO epochs on the same rollout batch by repeatedly re-evaluating $\ell_\theta^{(k)}$ under the frozen time/noise and forming the importance ratio
\begin{equation}
\rho^{(k)}(\theta)
=
\exp\!\left(\ell_{\theta_{\mathrm{old}}}^{(k)} - \ell_{\theta}^{(k)}\right).
\end{equation}
In practice, we clamp the log-ratio $\ell_{\theta_{\mathrm{old}}}^{(k)} - \ell_{\theta}^{(k)}$ to a fixed range before exponentiation for numerical stability.

We then optimize the clipped GRPO/PPO objective in Eq.~\eqref{eq:grpo}, using $A^{(k)}$ (single-objective, Eq.~\eqref{eq:grpo_adv}) or $A_{\lambda}^{(k)}$ (Eq.~\eqref{eq:grpo_adv}) as the advantage signal.

\textbf{KL regularization. }
\label{sec:kl_k3}
Following standard practice, we discourage the policy from drifting far from the pre-trained reference $\theta_{\mathrm{ref}}$, by adding a KL penalty computed on the rollout batch. We once again plug in the single-time surrogate described in Section~\ref{sec:single_time_surrogate},  using the same frozen time/noise to stay consistent between importance-ratio and KL-divergence computations. Defining
\begin{equation}
\Delta^{(k)} = \ell_\theta^{(k)} - \ell_{\theta_{\mathrm{ref}}}^{(k)},
\end{equation}
and using the surrogate identification $\ell \approx -\log \pi$, $\Delta^{(k)}$ approximates $\log\frac{\pi_{\mathrm{ref}}}{\pi_\theta}$.
After exponentiation, we then use the unbiased low-variance estimator as described by \citet{schulman2020klapprox}, and used in \cite{shao2024deepseekmath}:
\begin{equation}
\label{eq:k3}
\mathrm{KL}^{(k)} \;=\; \exp(\Delta^{(k)}) - \Delta^{(k)} - 1,
\end{equation}
and average across the group. The importance of KL in training stability is shown in Fig.~\ref{fig:Figure7}. 

There has been recent work by \citet{hoellmer2026open} that uses an alternative approach for inference-time RL by converting the ODE into an SDE via noise injection along the path. We instead keep deterministic ODE rollouts for efficiency and reduced hyperparameter burden: ODE samplers can reach comparable sample quality with fewer function evaluations than SDE-based samplers \cite{song2020score,song2020denoising}, and they avoid specifying an explicit stochastic noise scale and schedule (e.g., a $\sigma(t)$ schedule) during rollouts, which is particularly convenient when sampling is performed at every outer-loop iteration.

% There has been recent work by \citet{hoellmer2026open} that uses an alternative approach for inference-time RL by converting the ODE into an SDE by injecting noise into the path. 

\subsubsection{Advantage estimation}
\label{sec:advantage_estimation}
For a group of size $K$, we compute standard group relative advantages by standardizing rewards within the group and clipping:
\begin{equation}
\label{eq:grpo_adv}
A^{(k)}
=
\mathrm{clip}\!\left(
\frac{r^{(k)} - \mu_r}{\sigma_r + \epsilon_{\mathrm{adv}}},
\, -c, \, c
\right),
\end{equation}
where $\mu_r$ and $\sigma_r$ are the within-group mean and standard deviation, $\epsilon_{\mathrm{adv}}=10^{-8}$ is a numerical stabilizer, and $c$ is a fixed clipping threshold.

For multi-objective alignment, we introduce advantage mixing (Eq.~\eqref{eq:grpo_adv}), which (i) computes normalized, pre-clipped advantages separately for $r_E$ and $r_F$ within each group, (ii) mixes them using $\lambda\in[0,1]$, and (iii) applies the same clipping threshold $c$. This avoids manual reward rescaling when mixing objectives.

\subsubsection{Optimization objective}
We maximize the GRPO objective over multiple epochs:
\begin{equation}
\label{eq:grpo}
\mathcal{J}_{\mathrm{GRPO}}(\theta)
=
\frac{1}{K}\sum_{k=1}^K
\min\!\Big(
\rho^{(k)}(\theta)\,A^{(k)},\;
\mathrm{clip}(\rho^{(k)}(\theta),1-\epsilon,1+\epsilon)\,A^{(k)}
\Big)
\;-\;
\beta\cdot \frac{1}{K}\sum_{k=1}^K \mathrm{KL}^{(k)},
\end{equation}
where $\epsilon$ is the PPO clipping parameter and $\beta$ controls the strength of regularization to the frozen reference model. $\rho^{(k)}$ and $\mathrm{KL}^{(k)}$ are the importance-ratio and KL divergence respectively, which we approximate using single-time surrogate scores (Section~\ref{sec:single_time_surrogate}).

\subsection{Evaluation metrics}
\label{sec:eval_metrics}

We evaluate the quality of generated heavy-atom crystal proposals using the metrics reported in Tables~\ref{tab:Table1}--\ref{tab:Table2}. Throughout, a crystal is represented by heavy-atom types $\{a_i\}_{i=1}^{N}$, Cartesian coordinates $x$, and lattice parameters $\ell=(a,b,c,\alpha,\beta,\gamma)\in\mathbb{R}^6$ with corresponding lattice matrix $L(\ell)\in\mathbb{R}^{3\times 3}$ (\texttt{Pymatgen} convention). Unless stated otherwise, neighbor computations respect periodic boundary conditions.

\subsubsection{Density error (\%)}
For each structure, we compute the crystal density (in g/cm$^3$) from the unit-cell volume and total atomic mass:
\begin{align}
\rho(\ell, \{a_i\})
= \frac{m(\{a_i\})}{V(\ell)} \cdot 1.660539,
\qquad
V(\ell)=\left|\det L(\ell)\right|,
\end{align}
where $m(\{a_i\})=\sum_{i=1}^{N} m_{a_i}$ is the sum of element masses in atomic mass units (amu), and the constant $1.660539$ converts $\text{amu}/\text{\AA}^3$ to g/cm$^3$. We report the mean absolute percentage error (MAPE) between predicted and ground-truth densities:
\begin{align}
\mathrm{DensityError}(\%)=
100\cdot
\frac{1}{M}\sum_{n=1}^{M}
\left|
\frac{\rho(\ell^{(n)}_{\text{pred}},\{a_i^{(n)}\})-\rho(\ell^{(n)}_{\text{gt}},\{a_i^{(n)}\})}
{\rho(\ell^{(n)}_{\text{gt}},\{a_i^{(n)}\})+\varepsilon}
\right|,
\end{align}
with $\varepsilon=10^{-6}$ for numerical stability and $M$ the number of evaluated crystals.

\subsubsection{Clash rate (\%)}
We quantify unphysical short-range overlaps in a generated heavy-atom structure by detecting atoms whose closest periodic neighbors are too near relative to covalent radii.
Let $r(a_i)$ denote the covalent radius for element $a_i$. For each periodic neighbor pair $(i,j)$ within a cutoff, let $d_{ij}$ be the minimum-image distance under periodic boundary conditions. A pair is flagged as a \emph{clash} if
\begin{align}
d_{ij} < \alpha\big(r(a_i)+r(a_j)\big),
\end{align}
with $\alpha=0.75$ in all experiments. We then mark an atom as ``in clash'' if it participates in \emph{any} clashing pair, and report the percentage of atoms involved in at least one clash:
\begin{align}
\mathrm{Clash}(\%) = 100\cdot \frac{1}{N}\sum_{i=1}^{N} \ind\!\Big\{\exists j:\ d_{ij}<\alpha(r(a_i)+r(a_j))\Big\}.
\end{align}
This definition targets severe overlaps.

\subsubsection{AMD distance ($L_\infty$)}
To measure structural proximity to the experimental target while remaining robust to periodicity, we use the periodic average-minimum-distance (AMD) descriptor \cite{widdowson2020average, widdowson2022resolving}. \citet{widdowson2022resolving} have utilized this descriptor to successfully de-duplicate molecular crystals in the CSD demonstrating the descriptor's resolving power on our chemical space.
Given a periodic point set with atom types $\{a_i\}$, lattice $L$, and Cartesian positions $x$, the AMD descriptor produces a length-$K$ vector $\mathrm{AMD}_K(x,L)\in\mathbb{R}^K$ (with $K{=}100$ here) summarizing ordered neighbor distances under periodic boundary conditions.
We report the $L_\infty$ distance between predicted and ground-truth AMD vectors:
\begin{align}
\mathrm{AMD}_{L_\infty}
=
\left\|
\mathrm{AMD}_K(x_{\text{pred}},L_{\text{pred}})
-
\mathrm{AMD}_K(x_{\text{gt}},L_{\text{gt}})
\right\|_\infty.
\end{align}

\subsubsection{Radial distribution function (RDF) Wasserstein distance}
We compare packing statistics using the distribution of periodic interatomic distances up to a maximum radius $r_{\max}$ (10~\AA\ in our evaluation). For a given structure, we collect all periodic neighbor distances $\{d_{ij}\}$ within $[0,r_{\max}]$ and form a normalized histogram (probability mass function) on a shared grid with $B$ bins (100 bins):
\begin{align}
p_b = \frac{1}{Z}\sum_{(i,j)} \ind\!\{d_{ij}\in \text{bin } b\}, \qquad \sum_{b=1}^{B} p_b = 1,
\end{align}
and analogously $q_b$ for the ground-truth structure. We report the 1D Wasserstein-1 distance computed from the discrete cumulative distributions:
\begin{align}
\mathrm{RDFWass}
=
\Delta r \sum_{b=1}^{B} \left| \mathrm{CDF}_p(b)-\mathrm{CDF}_q(b)\right|,
\qquad
\mathrm{CDF}_p(b)=\sum_{t\le b} p_t,
\end{align}
where $\Delta r=r_{\max}/B$ is the bin width. We additionally report a \emph{short-range} variant, \emph{RDF Wass (Short)}, computed on the restricted range $[0,5]$~\AA\ (same bin width) and re-normalized within that range before computing the Wasserstein distance. The short and long-range variants are fitness measures of local and global packing statistics respectively. 

\subsubsection{Wall-clock generation time (s)}
We report the average wall-clock time per generated heavy-atom proposal, measured for the generative sampling step only (ODE integration from noise to $t{=}0$ for both coordinates and lattice). This timing excludes downstream post-processing such as hydrogen addition, hydrogen-only relaxation, and full crystal relaxation/ranking, and is averaged over multiple generations under fixed  hardware settings used for all methods in Table~\ref{tab:Table2}. A single NVIDIA V100 GPU is used for flow matching inference and $32$ CPU cores with MPI parallelization are used for Genarris inference.

\subsection{Dataset construction and splitting}
\label{sec:dataset_splitting}

\subsubsection{CSD data extraction and dataset construction} \label{sec:CCDC_preprocess}
We construct our homomolecular crystal dataset from the Cambridge Structural Database (CSD) using the CSD Python API, following the extraction pipeline developed by Galanakis and colleagues (see \cite{galanakis2024rapid} and associated software for full implementation details). Starting from CSD refcodes, we (i) enumerate \emph{refcode families} (refcodes sharing a common alphabetic prefix), (ii) filter entries to retain only \emph{homomolecular} crystals and only structures composed of the target element set
\begin{align}
\{ \mathrm{C,H,N,O,F,Cl,Br,I,P,S,B,Si,Se,As} \},
\end{align}
and (iii) restrict the dataset to a fixed set of common space groups
\begin{align}
\{\mathrm{P1, P\overline{1}, P2_1, C2, Pc, Cc, P2_1/m, C2/m, P2/c, P2_1/c, P2_1/n, C2/c,}
\nonumber\\
\mathrm{P2_12_12, P2_12_12_1, Pca2_1, Pna2_1, Pbcn, Pbca, Pnma, R\overline{3}, I4_1/a}\},
\end{align}
with asymmetric-unit multiplicities restricted to $Z' \in \{1,2,3,4,5\}$. For each retained crystal, we expand symmetry operations to obtain full unit-cell contents (i.e., all symmetry mates) prior to downstream processing. Within each refcode family, the pipeline further clusters members by packing similarity and selects a set of unique structures.

After CSD extraction, we further filter the dataset to structures with at most 250 atoms in the unit cell. This cap bounds training and sampling cost and avoids pathological memory scaling for large unit cells.

\subsubsection{Family-based dataset splits}
We construct train/validation/test splits using a family-based exclusion strategy to prevent closely related refcodes from leaking across splits. We define a refcode \emph{family} by removing trailing digits from the refcode (e.g., \texttt{DOLBIR07} and \texttt{DOLBIR08} share the family \texttt{DOLBIR}). After grouping all filtered entries by family, we shuffle families randomly and assign families to splits as indivisible units using a greedy allocation scheme. This guarantees that no refcode family appears in more than one split.

The resulting split sizes are:
$|\mathcal{D}_{\mathrm{train}}| = 83{,}299$, $|\mathcal{D}_{\mathrm{val}}| = 30{,}364$, and $|\mathcal{D}_{\mathrm{test}}| = 37{,}558$, with all splits satisfying the homomolecular/species/space-group/$Z'$ constraints above and the $\leq 250$ atom filter, and containing disjoint refcode families.

\subsection{Hydrogen addition and relaxation protocol}
\label{sec:mlip_hydrogenation_relax}

\subsubsection{MLIP choice and hydrogen addition}
All energy/force evaluations and relaxations are performed with a pre-trained machine-learned interatomic potential (MLIP) \texttt{uma-s-1p1} model \citep{wood2025family} with the \texttt{omc} task configuration \citep{gharakhanyan2026open}. Because our processed crystal structures are heavy-atom only, we add hydrogens with 3D coordinates using \texttt{AddHs(addCoords=True)} from the \texttt{RDKit} library.

\subsubsection{Crystal relaxation}
After hydrogenation, we perform an H-only relaxation under periodic boundary conditions, keeping all heavy atoms and lattice fixed and relaxing only hydrogen coordinates. Hydrogens are added using RDKit’s local geometry without explicit periodic-boundary awareness, so initial H placements can create strong clashes with neighboring molecules or periodic images. A short H-only relaxation (with heavy atoms and lattice fixed) removes these artifacts and yields a stable starting point for full relaxation. We run for a maximum of 25 optimizer steps with a convergence threshold of $f_\mathrm{max} < 0.05$ eV \AA{}$^{-1}$.
We then run a full relaxation (all atoms, and lattice) starting from the H-only-relaxed structure. We use 1000 relaxation steps, and analogous convergence threshold of $f_\mathrm{max} < 0.05$ eV \AA{}$^{-1}$. All relaxations use the \texttt{FIRE} optimizer implemented in \texttt{ASE} \cite{hjorth2017atomic}.

\subsubsection{Relative lattice energy calculation}
We compute relative lattice energies from MLIP energies as
\begin{equation}
E^{\mathrm{latt}} \;=\; \frac{E_{\mathrm{crys}} - E_{\mathrm{ref}}}{Z},
\end{equation}
where $E_{\mathrm{crys}}$ is the energy of the fully-relaxed crystal under consideration $E_{\mathrm{ref}}$ is the energy of the fully-relaxed reference crystal, and $Z$ is the number of molecules in the unit cell. In Fig.~\ref{fig:Figure5}, the reference crystal is consistently chosen as the crystal for the refcode under consideration, having the global minimum relaxed energy.

\subsection{Baseline methods}
\label{sec:baselines}

We compare our method against two baselines based on the open-source \texttt{Genarris} package for molecular crystal structure generation: \emph{Genarris Plain} (v2.0) \cite{tom2020genarris} and \emph{Genarris Rigid Press} (v3.0) \cite{yang2025genarris}. Genarris Rigid Press has been used recently to successfully identify polymorphs of several rigid crystals by \citet{gharakhanyan2025fastcsp}, hence serving as a strong baseline method. We focus comparisons on baselines with publicly available implementations at the time of writing; this excludes OXtal from runnable benchmarking.

\subsubsection{Genarris Plain and Rigid Press}
\emph{Genarris Plain} generates random crystal structures by placing molecules into space groups using geometric constraints to prevent unphysical atomic overlaps, without performing energy evaluations. \emph{Genarris Rigid Press} extends this by applying a ``Rigid Press'' optimization algorithm, which uses a regularized hard-sphere potential to isotropically compress the unit cell. This achieves maximally close-packed structures based purely on geometric considerations, acting as a low-cost proxy for energetic stability.

For both baselines, we obtain the initial 3D conformer for the crystal by converting the reference SMILES string into a 3D conformer using RDKit (ETKDG method) \cite{landrum2013rdkit}. For more details on the exact configuration used for Genarris sampling, see Section~\ref{sec:genarris}.

\section{Code availability}
The corresponding code will be available at \href{https://github.com/learningmatter-mit/packflow}{https://github.com/learningmatter-mit/packflow}.

\section{Author contributions}
R.G.-B. and A.S. conceived the project. A.S. and E.P. were core contributors. A.S. led the research and designed the \texttt{PackFlow} architecture, pre-training method, baselines, and metrics with technical input from J.N., M.W., S.Q., and C.W.P. E.P. developed the post-training RL method with technical input from A.S. A.S. performed all evaluations and physics-based calculations with technical input from J.N. and E.P. A.S. and E.P. led the analysis of all results. A.S., E.P., J.N., and M.W. wrote the initial version of the manuscript; all authors reviewed and edited the final version. R.G.-B., E.O., and T.S.J. supervised the work.

\section{Acknowledgements}
A.S. acknowledges financial support from Sony Corporation, E.P. acknowledges funding from the Office of Naval Research (ONR) under contract N00014-20-1-2280, and J.N. acknowledges support from the Mathworks Fellowship.  We acknowledge the MIT SuperCloud and Lincoln Laboratory Supercomputing Center for providing high-performance computing resources. 
We also thank Max Baranov for feedback on the paper, and Max Baranov, Miguel Steiner, Mingrou Xie, Malte Franke, Nofit Segal, and members of the Learning Matter group at MIT for helpful suggestions and discussions throughout this work.

%%===========================================================================================%%
%% If you are submitting to one of the Nature Portfolio journals, using the eJP submission   %%
%% system, please include the references within the manuscript file itself. You may do this  %%
%% by copying the reference list from your .bbl file, paste it into the main manuscript .tex %%
%% file, and delete the associated \verb+\bibliography+ commands.                            %%
%%===========================================================================================%%

\bibliography{sn-bibliography}% common bib file
%% if required, the content of .bbl file can be included here once bbl is generated
%%\input sn-article.bbl

\begin{appendices}
\counterwithin{figure}{section}

\pagebreak

\setcounter{page}{1}
% \resetlinenumber

\begin{center}

{\LARGE Supplementary Information for}

{\LARGE \texttt{PackFlow}: Generative Molecular Crystal Structure Prediction via Reinforcement Learning Alignment}

\vspace{0.5cm}

{\large Akshay Subramanian$^{1}$, Elton Pan$^{1}$, Juno Nam$^1$, Maurice Weiler$^2$,
Shuhui Qu$^3$, Cheol Woo Park$^3$, Tommi S. Jaakkola$^2$, Elsa Olivetti$^1$,
Rafael Gomez-Bombarelli$^{1*}$}

\vspace{0.5cm}

$^1$ Department of Materials Science and Engineering, Massachusetts Institute of Technology, Cambridge, MA, 02139, USA.

$^2$ Electrical Engineering \& Computer Science Department, Massachusetts Institute of Technology, Cambridge, MA, 02139, USA.

$^3$ Samsung Display America Lab, San Jose, CA, 95134, USA.

\vspace{0.3cm}

$^*$ Correspondence to: rafagb@mit.edu

\end{center}

\begin{algorithm}[t]
\caption{Physics alignment (PA) of \texttt{PackFlow-Base} with GRPO and single-time surrogate}
\label{alg:grpo_packflow}
\begin{algorithmic}[1]
\Require pre-trained policy $v_\theta$ (\texttt{PackFlow-Base}) and frozen reference $v_{\theta_{\mathrm{ref}}}$,
dataset $\mathcal{D}$, group size $K$, PPO epochs $M$, clip $\epsilon$, KL coef $\beta$,
mixing weight $\lambda\in[0,1]$.
\For{each iteration}
  \State Sample templates $\{\tau_j\}_{j=1}^{B}\sim\mathcal{D}$
  \Statex \textbf{(A) Rollouts + rewards (no grad through sampling/MLIP)}
  \For{each $\tau_j$}
    \State $\{(x_0^{(k)},\ell_0^{(k)})\}_{k=1}^{K} \leftarrow \mathrm{SampleODE}(v_\theta,\tau_j)$
    \State $r_E^{(k)}\leftarrow -E_h^{(k)}$, \quad $r_F^{(k)}\leftarrow -F_h^{(k)}$
    \State $\{A^{(k)}\}_{k=1}^{K}\leftarrow \mathrm{AdvantageMixing}(\{r_E^{(k)}\},\{r_F^{(k)}\};\lambda)$
  \EndFor
  \State $\theta_{\mathrm{old}} \leftarrow \theta$ \Comment{policy that generated the rollouts}

  \Statex \textbf{(B) Freeze single-time noise/time and cache old scores}
  \For{each proposal $k$ in rollout batch}
    \State Draw $t^{(k)}\sim\mathcal{U}(0,1)$ and $(\varepsilon_x^{(k)},\varepsilon_\ell^{(k)})\sim\mathcal{N}(0,I)$; set $t_x^{(k)}=t_\ell^{(k)}\equiv t^{(k)}$
    \State $\ell_{\mathrm{old}}^{(k)} \leftarrow \mathrm{SingleTimeLoss}(\theta_{\mathrm{old}};\ x_0^{(k)},\ell_0^{(k)},t^{(k)},\varepsilon_x^{(k)},\varepsilon_\ell^{(k)})$
    \State $\ell_{\mathrm{ref}}^{(k)} \leftarrow \mathrm{SingleTimeLoss}(\theta_{\mathrm{ref}};\ x_0^{(k)},\ell_0^{(k)},t^{(k)},\varepsilon_x^{(k)},\varepsilon_\ell^{(k)})$
  \EndFor

  \Statex \textbf{(C) PPO-style GRPO updates (reuse frozen $(t,\varepsilon)$)}
  \For{$m=1$ to $M$}
    \State $\hat{\mathcal{J}}\leftarrow 0$
    \For{each proposal $k$ in rollout batch}
      \State $\ell^{(k)} \leftarrow \mathrm{SingleTimeLoss}(\theta;\ x_0^{(k)},\ell_0^{(k)},t^{(k)},\varepsilon_x^{(k)},\varepsilon_\ell^{(k)})$
      \State $\rho^{(k)} \leftarrow \exp\!\big(\mathrm{clip}(\ell_{\mathrm{old}}^{(k)}-\ell^{(k)},-d,+d)\big)$
      \State $\Delta^{(k)} \leftarrow \ell^{(k)}-\ell_{\mathrm{ref}}^{(k)}$
      \State $\mathrm{KL}^{(k)} \leftarrow \exp(\Delta^{(k)})-\Delta^{(k)}-1$ \Comment{Eq.~\eqref{eq:k3}}
      \State $g^{(k)} \leftarrow \min\!\Big(\rho^{(k)}A^{(k)},\ \mathrm{clip}(\rho^{(k)},1-\epsilon,1+\epsilon)A^{(k)}\Big)$
      \State $\hat{\mathcal{J}}\leftarrow \hat{\mathcal{J}} + \frac{1}{K}g^{(k)} - \beta\cdot \frac{1}{K}\mathrm{KL}^{(k)}$
    \EndFor
    \State $\theta \leftarrow \theta + \eta \nabla_\theta \hat{\mathcal{J}}$
  \EndFor
\EndFor
\end{algorithmic}
\end{algorithm}

\begin{algorithm}[t]
\caption{Advantage Mixing: mix \emph{normalized} (pre-clipped) group relative advantages}
\label{alg:grpo_adv}
\begin{algorithmic}[1]
\Function{AdvantageMixing}{$\{r_E^{(k)}\}_{k=1}^K,\{r_F^{(k)}\}_{k=1}^K;\lambda,c,\epsilon_{\mathrm{adv}}$}
  \State Compute within-group mean/std for energy rewards:
  \[
    \mu_E \leftarrow \frac{1}{K}\sum_{k=1}^K r_E^{(k)},
    \quad
    \sigma_E \leftarrow \sqrt{\frac{1}{K}\sum_{k=1}^K (r_E^{(k)}-\mu_E)^2} + \epsilon_{\mathrm{adv}}
  \]
  \State Compute within-group mean/std for force rewards:
  \[
    \mu_F \leftarrow \frac{1}{K}\sum_{k=1}^K r_F^{(k)},
    \quad
    \sigma_F \leftarrow \sqrt{\frac{1}{K}\sum_{k=1}^K (r_F^{(k)}-\mu_F)^2} + \epsilon_{\mathrm{adv}}
  \]
  \State Compute \emph{normalized, pre-clipped} advantages separately:
  \[
    \tilde A_E^{(k)} \leftarrow \frac{r_E^{(k)}-\mu_E}{\sigma_E},\qquad
    \tilde A_F^{(k)} \leftarrow \frac{r_F^{(k)}-\mu_F}{\sigma_F}
  \]
  \State Mix normalized advantages, then apply a shared clipping threshold:
  \[
    A_\lambda^{(k)} \leftarrow \lambda\,\tilde A_E^{(k)} + (1-\lambda)\,\tilde A_F^{(k)},
    \qquad
    A^{(k)} \leftarrow \mathrm{clip}(A_\lambda^{(k)},-c,+c)
  \]
  \State \Return $\{A^{(k)}\}_{k=1}^K$
\EndFunction
\end{algorithmic}
\end{algorithm}

\begin{algorithm}[t]
\caption{Single-time surrogate loss (flow-matching integrand proxy for $-\log\pi_\theta$)}
\label{alg:single_time_loss}
\begin{algorithmic}[1]
\Function{SingleTimeLoss}{$\theta;\ x_0,\ell_0,t,\varepsilon_x,\varepsilon_\ell$}
  \State Form OT interpolation noisy states and targets (shared time $t$):
  \[
    x_t \leftarrow (1-t)\,x_0 + t\,\varepsilon_x,
    \qquad
    \ell_t \leftarrow (1-t)\,\ell_0 + t\,\varepsilon_\ell
  \]
  \[
    u_x \leftarrow \varepsilon_x - x_0,
    \qquad
    u_\ell \leftarrow \varepsilon_\ell - \ell_0
  \]
  \State Predict velocity fields with \texttt{PackFlow}:
  \[
    (v_{x,\theta},v_{\ell,\theta}) \leftarrow v_\theta(x_t,\ell_t,t,t)
  \]
  \State Return per-sample surrogate loss (Eq.~(single-time loss)):
  \[
    \ell_\theta \leftarrow
    \frac{1}{N}\left\lVert v_{x,\theta}-u_x\right\rVert_2^2
    + \lambda_\ell\left\lVert v_{\ell,\theta}-u_\ell\right\rVert_2^2
  \]
  \State \Return $\ell_\theta$
\EndFunction
\end{algorithmic}
\end{algorithm}

\begin{table}[t]
\centering
\small
\begin{tabularx}{0.5\linewidth}{@{}l l@{}}
\toprule
\textbf{Hyperparameter} & \textbf{Value} \\
\midrule

% \multicolumn{2}{@{}l}

% {\textbf{Data}}\\
% \texttt{data\_dir} & \texttt{/home/gridsan/sakshay/experiments/packflow/processed\_data\_lt200\_symmetrized\_niggli\_tag\_hbond\_hydrogen\_aromatic\_rings} \\
% \addlinespace

\multicolumn{2}{@{}l}{\textbf{Objective / Representation}}\\
\texttt{objective} & \texttt{flow\_matching} \\
\texttt{coordinate\_system} & \texttt{cartesian} \\
\texttt{lattice\_loss\_weight} & \texttt{1.0} \\
\addlinespace

\multicolumn{2}{@{}l}{\textbf{Model}}\\
\texttt{model} & \texttt{PackFlow-Base-60M} \\
\texttt{d\_model} & \texttt{640} \\
\texttt{nhead} & \texttt{10} \\
\texttt{dim\_feedforward} & \texttt{2560} \\
\texttt{num\_transformer\_layers} & \texttt{12} \\
\texttt{disable\_positional\_embeddings} & \texttt{True} \\
\addlinespace

\multicolumn{2}{@{}l}{\textbf{Graph conditioning}}\\
\texttt{use\_rdkit\_features} & \texttt{True} \\
\texttt{use\_attention\_bias\_from\_graph} & \texttt{True} \\
\addlinespace

\multicolumn{2}{@{}l}{\textbf{Training}}\\
\texttt{n\_epochs} & \texttt{10000} \\
\texttt{batch\_size} & \texttt{128} \\
\texttt{learning\_rate} & \texttt{3e-5} \\
\texttt{optimizer} & \texttt{AdamW} \\
\texttt{max\_grad\_norm} & \texttt{1.0} \\
\addlinespace

\multicolumn{2}{@{}l}{\textbf{Distributed / Logging}}\\
\texttt{distributed} & \texttt{True (DDP)} \\
\texttt{nodes} & \texttt{4} \\
\texttt{gpus\_per\_node} & \texttt{2} \\
% \texttt{use\_wandb} & \texttt{True} \\
% \texttt{wandb\_mode} & \texttt{offline} \\
\addlinespace

\bottomrule
\end{tabularx}
\caption{Base hyperparameters for \texttt{PackFlow} pre-training (Base-60M).}
\label{tab:pretrain_hyperparams}
\end{table}

\begin{table}[t]
\centering
\small
\begin{tabularx}{0.5\linewidth}{@{}l l@{}}
\toprule
\textbf{Hyperparameter} & \textbf{Value} \\
\midrule

\multicolumn{2}{@{}l}{\textbf{Data}}\\
\texttt{max\_train\_samples} & \texttt{8000} \\
\addlinespace

\multicolumn{2}{@{}l}{\textbf{Sampling}}\\
\texttt{num\_seeds} & \texttt{8} \\
\texttt{n\_steps} & \texttt{200} \\
\texttt{lambda\_val} & \texttt{1.0} \\
\addlinespace

\multicolumn{2}{@{}l}{\textbf{Rewards / Advantages}}\\
\texttt{reward\_type} & \texttt{ef} \\
\texttt{ef\_lambda} & \texttt{0.0 / 0.5 / 1.0} \\
\texttt{advantage\_eps} & \texttt{1e-8} \\
\texttt{advantage\_clip} & \texttt{5.0} \\
\addlinespace

\multicolumn{2}{@{}l}{\textbf{GRPO/PPO}}\\
\texttt{ppo\_epochs} & \texttt{4} \\
\texttt{ppo\_clip\_range} & \texttt{0.2} \\
\texttt{kl\_coef} & \texttt{1} \\
\addlinespace

\multicolumn{2}{@{}l}{\textbf{Training}}\\
\texttt{epochs} & \texttt{100} \\
\texttt{batch\_size} & \texttt{16} \\
\texttt{learning\_rate} & \texttt{3e-7} \\
\texttt{weight\_decay} & \texttt{0.0} \\
\texttt{max\_grad\_norm} & \texttt{1.0} \\
\texttt{warmup\_epochs} & \texttt{0.0} \\
\texttt{coords\_loss\_weight} & \texttt{1.0} \\
\texttt{lattice\_loss\_weight} & \texttt{1.0} \\
\addlinespace

\bottomrule
\end{tabularx}
\caption{Base hyperparameters for physics alignment post-training.}
\label{tab:PA_hyperparams}
\end{table}

\section{Validity of heavy-atom proxy rewards}
\label{sec:proxy_rewards}

Physics alignment uses heavy-atom energies and forces computed before hydrogen addition and full relaxation as a tractable proxy for the downstream stability signals that would be obtained after hydrogen completion and hydrogen-only relaxation. This approximation reduces the cost of each GRPO update by avoiding per-sample hydrogen placement and relaxation inside the post-training loop, while still providing dense feedback about steric overlap and local strain (via force-based signals) and energetic favorability (via energies). Because GRPO updates depend on \emph{group relative advantages} rather than absolute reward scales (Section~\ref{sec:rl_methods}), the key question is whether proxy-based rewards induce \emph{similar within-group advantage signals} to those obtained from corresponding with-H evaluations.

\subsection{Proxy versus with-H advantage signals}
For each crystal template, we sample a group of $K$ heavy-atom proposals and evaluate two reward channels:
(i) an energy-based reward $r_E^{(k)}=-E^{(k)}$ and
(ii) a force-based reward $r_F^{(k)}=-F^{(k)}$,
where $E^{(k)}$ and $F^{(k)}$ are computed either on the heavy-atom (no-H) structure or on the hydrogenated (with-H) structure after hydrogen addition.

From these rewards, we compute the GRPO learning signals using the \emph{same} group-standardization and clipping rules used during training: for single-objective alignment we use the group relative advantage in Eq.~\eqref{eq:grpo_adv}; for multi-objective alignment we use the mixed advantage in Eq.~\eqref{eq:grpo_adv}. This yields two $K$-dimensional vectors per group:
$A_{\mathrm{noH}}^{(\lambda)} \in \mathbb{R}^K$ from heavy-atom proxy evaluations, and
$A_{\mathrm{withH}}^{(\lambda)} \in \mathbb{R}^K$ from with-H evaluations, where $\lambda\in\{0,0.5,1\}$ matches the post-training modes considered in Fig.~\ref{fig:Figure4}.

\subsection{Alignment metrics}
We quantify agreement between proxy- and with-H-based learning signals using two statistics computed per group and then averaged across groups:
(i) \emph{advantage vector alignment} (AVA), the cosine similarity
\begin{align}
\mathrm{AVA}
\;=\;
\frac{\langle A^{(\lambda)}_{\mathrm{noH}},\,A^{(\lambda)}_{\mathrm{withH}}\rangle}
{\|A^{(\lambda)}_{\mathrm{noH}}\|_2\,\|A^{(\lambda)}_{\mathrm{withH}}\|_2},
\end{align}
and (ii) \emph{sign agreement}, the fraction of samples whose advantages share the same sign:
\begin{align}
\mathrm{SignAgree}
\;=\;
\frac{1}{K}\sum_{k=1}^{K}
\mathbb{I}\!\left[\mathrm{sign}\!\left(A^{(\lambda)}_{\mathrm{noH},k}\right)
=
\mathrm{sign}\!\left(A^{(\lambda)}_{\mathrm{withH},k}\right)\right].
\end{align}
High AVA indicates that proxy- and with-H-based GRPO would reweight samples in similar directions within the group, while high sign agreement indicates consistent ``good/bad'' assignments across proposals.

\subsection{Empirical proxy fidelity}
Across 90 crystals with average group size $K=20$, heavy-atom proxy rewards yield strongly aligned advantage signals with the corresponding with-H advantages for all three post-training modes:
\begin{align}
\lambda=1\ :\;& \mathrm{AVA}=0.793,\;\; \mathrm{SignAgree}=0.813; \\
\lambda=0\ :\;& \mathrm{AVA}=0.908,\;\; \mathrm{SignAgree}=0.883; \\
\lambda=0.5\ :\;& \mathrm{AVA}=0.872,\;\; \mathrm{SignAgree}=0.855.
\end{align}
These values indicate that, after applying the \emph{same} GRPO standardization and clipping used during training (Eqs.~\eqref{eq:grpo_adv}--\eqref{eq:grpo_adv}), heavy-atom energy/force rewards produce within-group learning signals that closely track their with-H counterparts. Together, these results support the use of heavy-atom energies and forces as a computationally efficient yet informative proxy for physics alignment.

\section{Canonical unwrapping of molecules}
\label{sec:unwrapping}
Standard crystallographic data formats typically constrain fractional coordinates to the unit interval $[0, 1)^3$, which frequently results in ``fragmented'' molecules where covalently bonded atoms are wrapped to opposite faces of the unit cell. To recover physically valid molecular geometries for the generative model, we implement a graph-based unwrapping procedure. We first isolate connected components within the crystal graph based on covalent connectivity. For each component, we reconstruct the continuous molecule using a Breadth-First Search (BFS) traversal: for every bonded pair of atoms $(i, j)$ visited, we apply an integer lattice shift $n \in \{-1, 0, 1\}^3$ to the coordinates of atom $j$ that minimizes the distance to the already-visited atom $i$ (the minimum image convention).To ensure a deterministic and canonical representation invariant to the input atom ordering, we employ a two-pass reconstruction strategy. The first pass resolves the gross topology to calculate the molecule's geometric centroid. We then identify the ``central atom'', defined as the atomic site closest to this centroid, and execute a second BFS traversal rooted specifically at this atom. This centering step minimizes arbitrary spatial drift and ensures that the unwrapped coordinates are consistently defined relative to the molecule's geometric center. This unwrapping procedure is visually depicted in Fig.~\ref{fig:Figure2}b.

\section{Symmetries and gauge-fixing in the data representation} \label{sec:symmetries}
A periodic crystal admits multiple coordinate descriptions that correspond to the same physical structure.
Rather than enforcing that \texttt{PackFlow} is invariant to these symmetries, we \emph{fix a single representative} of each crystal during preprocessing and train the model on that canonicalized representation (Fig.~\ref{fig:Figure2}b).
Concretely, we first \emph{unwrap} each molecule across periodic boundaries using covalent connectivity before converting to Cartesian coordinates.
This avoids discontinuities that arise when bonded atoms are wrapped to opposite faces of the unit cell (i.e., bond vectors differing by integer lattice translations), yielding smoother bond geometry and pairwise-distance statistics for the model to learn. This unwrapping is performed so that a side for unwrapping is chosen canonically to be the side that contains the molecular centroid, allowing the majority of each molecule to always lie inside the unit cell. More details on the exact algorithm used for this preprocessing are provided in Section~\ref{sec:unwrapping}. Using an unwrapped representation was an essential design choice, since our initial attempts with fractional coordinates and wrapped cartesian coordinates (similar to recent inorganic CSP methods such as \cite{miller2024flowmm}) were unable to train effectively.

\subsection{Global translation}
After unwrapping, we remove the remaining global translation degree of freedom by mean-centering,
$\tilde{x} = x - \frac{1}{N}\mathbf{1}_N \mathbf{1}_N^\top x$,
so the model is trained and sampled in a centered coordinate frame.

\subsection{Rotation}
A global rotation acts as $\mathbf{x}_i \mapsto R\,\mathbf{x}_i$ for each atom position vector $\mathbf{x}_i \in \mathbb{R}^3$, and $R\in SO(3)$.
Equivalently, if coordinates are stored as a matrix $x\in\mathbb{R}^{N\times 3}$ with rows $\mathbf{x}_i^\top$, then $x \mapsto xR^\top$. Because we do not apply rotational augmentation and the architecture is not explicitly rotation-equivariant, \texttt{PackFlow}-generated distributions are not $SO(3)$ invariant.
Instead, we express all lattices using a deterministic lattice-parameter-to-matrix convention (\texttt{Pymatgen}; $c\parallel z$ and a right-handed cell embedding), which removes arbitrary frame-to-frame rotations in the stored representation and improves sample efficiency.

\section{Genarris sampling and shortlisting strategy} \label{sec:genarris}
To ensure a fair comparison with a fixed sampling budget, we configure Genarris to generate a candidate pool large enough to guarantee diversity. Our sampling config is shown in Listing~\ref{lst:genarris_config}. We set the internal generation parameter \texttt{num\_structures\_per\_spg} equal to the target number of samples, $K$. Genarris attempts to generate this many valid structures for each space group in its standard search distribution.

From the resulting pool of valid raw candidates, which typically exceeds $K$, we shortlist the final set of $K$ proposals using a space-group-stratified sampling strategy. We identify all unique space groups $S$ represented in the candidate pool and assign a selection quota $q_s \approx K / |S|$ to each space group. We randomly sample $q_s$ unique structures from each space group's candidates. Any remaining deficit required to reach exactly $K$ samples is filled by random sampling from the remaining pool of unselected valid structures, ensuring the final baseline set is both size-constrained and structurally diverse.

\begin{lstlisting}[basicstyle=\ttfamily\small, frame=single, caption={Genarris configuration used for Plain and Rigid Press baselines.}, label={lst:genarris_config}]
[master]
name = genarris_run
molecule_path = ["molecule.xyz"]
Z = <Z_VALUE>
log_level = info

[workflow]
tasks = ['generation']  # Plain
# tasks = ['generation', 'symm_rigid_press']  # Rigid Press

[generation]
# Budget per space group.
num_structures_per_spg = <max(10, N)>
sr = 0.85
max_attempts_per_spg = 100000
tol = 0.1
unit_cell_volume_mean = predict
volume_mult = 1.0
max_attempts_per_volume = 1000
spg_distribution_type = standard
generation_type = crystal
natural_cutoff_mult = 1.0

[symm_rigid_press]  # Only used for the Rigid Press baseline
sr = 0.85
method = BFGS
tol = 0.01
natural_cutoff_mult = 1.0
debug_flag = False
maxiter = 100
\end{lstlisting}

\noindent
Here, \texttt{<Z\_VALUE>} is inferred per target crystal (unit-cell multiplicity), and \texttt{N} is the number of final baseline samples we require per target (e.g., \texttt{N = num\_samples} in our evaluation pipeline).

\section{Sample efficiency}\label{secA1}
\begin{figure}[h!]
\centering
\includegraphics[width=0.6\textwidth]{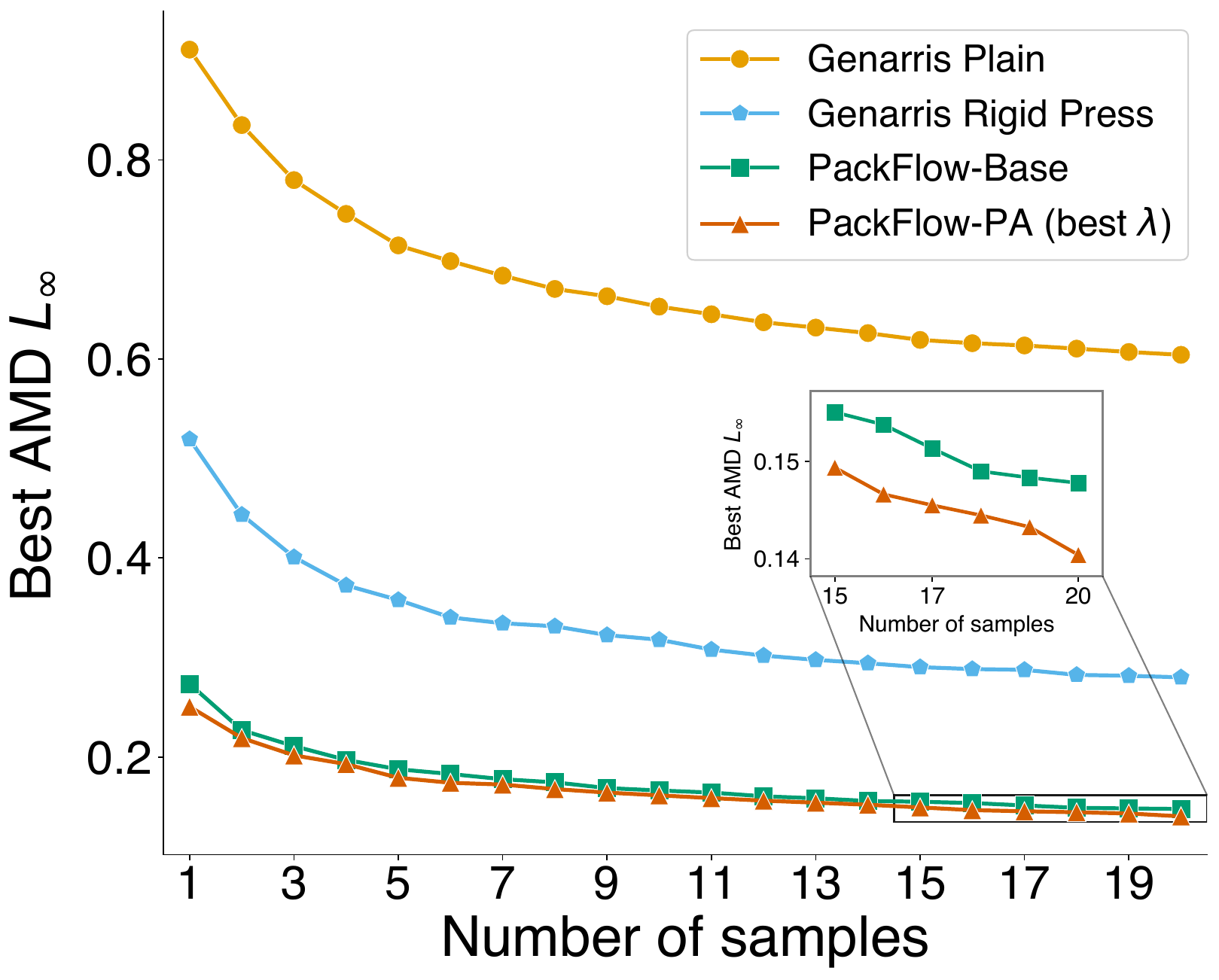}
\caption{\textbf{Sample efficiency of \texttt{PackFlow} and Genarris predictions. }  \texttt{PackFlow} models reach closer to experimental structures at a significantly faster rate than Genarris baselines. The lines represent mean AMD $L_{\infty}$ scores over the entire test set. }
\label{fig:Figure6}
\end{figure}

\texttt{PackFlow} models reach significantly lower AMD $L_{\infty}$ than Genarris baselines at faster rates (Fig.~\ref{fig:Figure6}) when averaged across the entire test set.

\section{Stability of post-training}\label{secA2}
\begin{figure}[ht!]
\centering
\makebox[\linewidth][c]{%
  \includegraphics[width=1.25\linewidth]{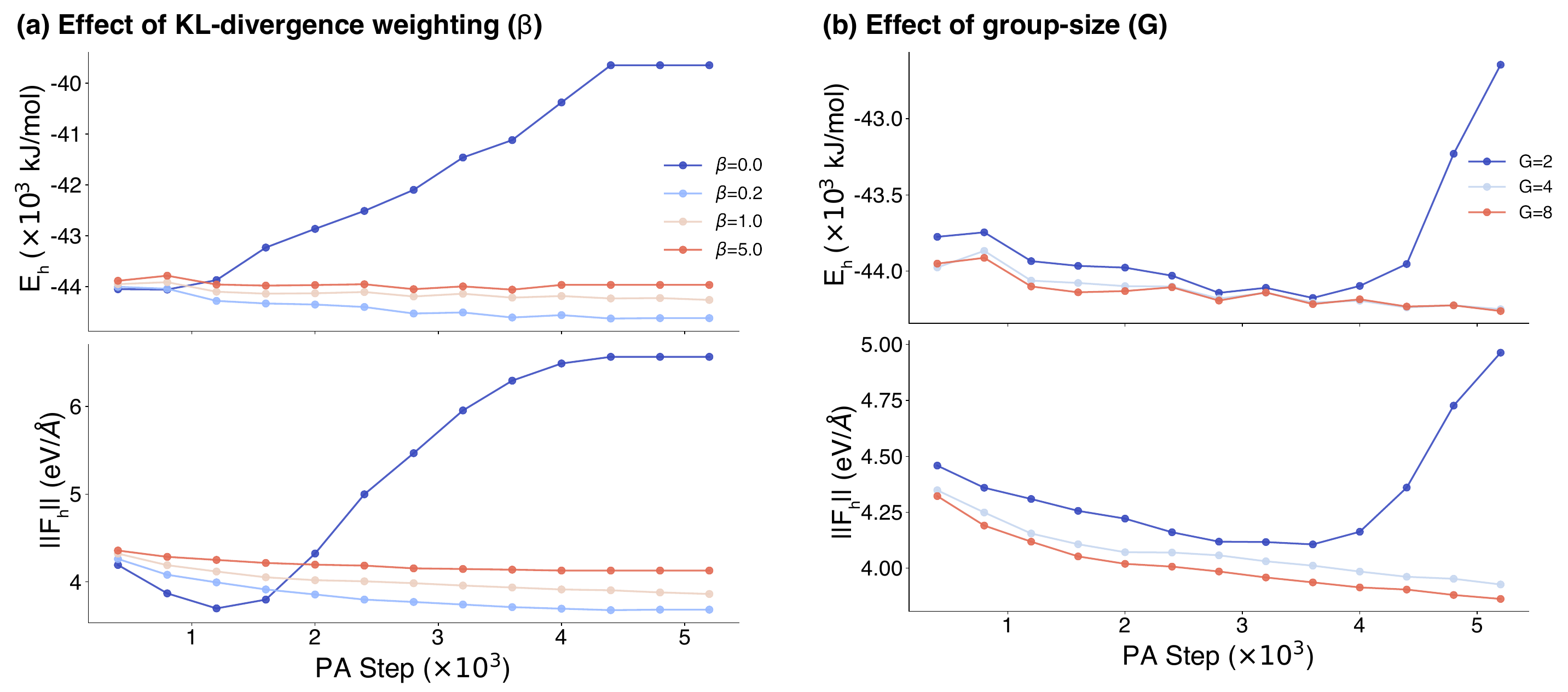}%
}
\caption{\textbf{Effects of KL-divergence weighting $\beta$ and group-size $G$ on training stability during physics alignment (PA). }  Lack of KL-divergence, and small group-size results in instability during post-training demonstrated in \textbf{(a)} and \textbf{(b)} through large spikes in values of heavy-atom energies and forces during training. }
\label{fig:Figure7}
\end{figure}

Fig.~\ref{fig:Figure7}a emphasizes the importance of having non-zero KL contribution in our task; $\beta = 0$ results in significant training instability in the heavy-atom energies ($E_h$) and forces ($F_h$). This is in contrast to some prior works where exclusion of the KL term has been an intentional design choice \cite{yu2025dapo}. In similar spirit, we also observed that small group-sizes lead to similar training instability (Fig.~\ref{fig:Figure7}b). Larger group sizes lead to more well-behaved training.

% \begin{figure}[h!]
% \centering
% \includegraphics[width=1\textwidth]{figures/Figure7.pdf}

% ===================== SI ADDITION: FIGURE 5 VISUALIZATION FILTERS =====================

\section{Visualization filters for Fig.~\ref{fig:Figure5}}
\label{sec:fig5_vis_filters}

For the density--energy visualizations in Fig.~\ref{fig:Figure5}, we apply a lightweight
\emph{bond-length validity} filter to remove relaxed structures that exhibit clearly unphysical
\emph{intramolecular} geometries (e.g., broken covalent bonds or severely over-compressed bonds).
This filter is used only for producing clean visual summaries and does not change the underlying
generation or relaxation procedure.

\subsection{Bond-length validity (covalent radii) filter}
\label{sec:bond_length_filter}

Each generated crystal comes with an intramolecular covalent bond list (heavy-atom graph edges)
$\mathcal{E}$, provided by the template connectivity.
For each bond $(i,j)\in\mathcal{E}$, we compute the minimum-image distance under periodic boundary
conditions (PBC),
\begin{equation}
d_{ij}
\;=\;
\left\lVert
\big( (x_j - x_i)L^{-1} - \mathrm{round}((x_j - x_i)L^{-1}) \big)L
\right\rVert_2,
\end{equation}
where $x_i,x_j\in\mathbb{R}^3$ are Cartesian coordinates and $L\in\mathbb{R}^{3\times 3}$ is the lattice matrix.

We define an expected covalent bond length using tabulated covalent radii $r(\cdot)$:
\begin{equation}
d_{ij}^{\mathrm{exp}} = r(a_i) + r(a_j),
\end{equation}
where $a_i$ and $a_j$ are the element types of atoms $i$ and $j$.
A bond is considered valid if its PBC distance lies within a symmetric tolerance window,
\begin{equation}
\big|d_{ij} - d_{ij}^{\mathrm{exp}}\big| \le \delta,
\qquad \delta = 0.4~\text{\AA}.
\label{eq:bond_length_filter}
\end{equation}
A structure is marked \emph{bond-length valid} if \emph{all} bonds satisfy Eq.~\eqref{eq:bond_length_filter}.
In practice, this filter removes structures where the relaxation has produced an implausible molecular geometry
(e.g., bond breaking/formation artifacts) that would otherwise confound qualitative comparisons in Fig.~\ref{fig:Figure5}.

\section{Comparison of Base and PA models}
\label{sec:basevspa}

% \begin{figure}[h!]
% \centering
% \includegraphics[width=1\textwidth]{figures/Figure8.pdf}
\begin{figure}[ht!]
\centering
\makebox[\linewidth][c]{%
  \includegraphics[width=1.25\linewidth]{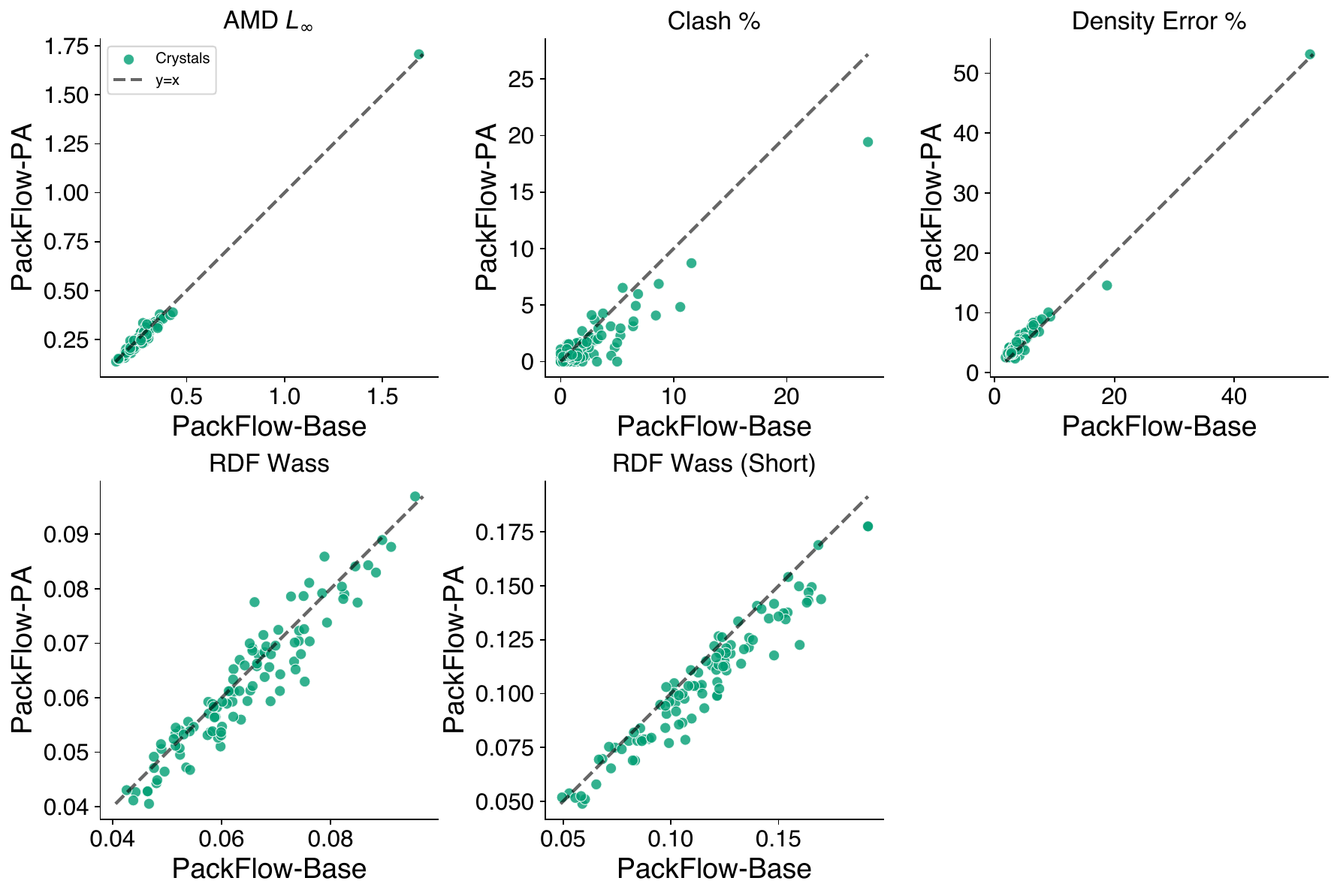}%
}
\caption{\textbf{Comparison of base and PA models.}  It is clear from Clash \% and RDF Wass (Short) scatter plots that physics alignment shifts the points towards better values, indicating that the primary effects of physics alignment are in resolving short-range unphysical contacts.}
\label{fig:Figure8}
\end{figure}

\section{Atom and bond features}
\subsection{RDKit atom and bond features}
\label{sec:rdkit_features}

To condition \texttt{PackFlow} on molecular identity and connectivity, we encode each molecule as a graph using \texttt{RDKit}. Each heavy atom $i$ is assigned a discrete feature vector $f_i$ (node features), and each covalent bond $(i,j)$ is assigned a discrete edge feature vector $b_{ij}$ (bond features). We use only the atom/bond attributes shown in our implementation (\texttt{featurize\_atoms} and \texttt{featurize\_bond}).

\subsubsection{Discrete feature encoding}
All scalar/categorical features are converted into integer indices using a fixed vocabulary (Python lists \texttt{atom\_features\_list} and \texttt{bond\_features\_list}). We apply a \texttt{safe\_index} mapping: if a value is not present in the corresponding vocabulary, it is mapped to the final ``misc'' bucket (the last entry of the list). Boolean features use the vocabulary \texttt{[False, True]}.

\subsubsection{Atom features (\texttt{featurize\_atoms})}
Given an \texttt{RDKit} molecule \texttt{mol}, we compute ring metadata via \texttt{ringinfo = mol.GetRingInfo()}. We call \texttt{atom.UpdatePropertyCache()} for all atoms prior to feature extraction to ensure valence-related properties are up-to-date. Table~\ref{tab:rdkit_atom_features} summarizes the per-atom features used.

\begin{table}[t]
\centering
\small
\setlength{\tabcolsep}{7pt}
\renewcommand{\arraystretch}{1.15}
\caption{RDKit-derived atom features used in \texttt{featurize\_atoms}. Each feature is mapped to an integer index via \texttt{safe\_index} against a fixed vocabulary.}
\label{tab:rdkit_atom_features}
\begin{tabular}{p{0.26\linewidth} p{0.40\linewidth} p{0.26\linewidth}}
\toprule
\textbf{Feature key} & \textbf{RDKit source} & \textbf{Description} \\
\midrule
\texttt{atomic\_num} &
\texttt{atom.GetAtomicNum()} &
Atomic number (element identity). \\

\texttt{degree} &
\texttt{atom.GetTotalDegree()} &
Total degree (number of directly bonded neighbors, incl.\ H as configured by RDKit). \\

\texttt{numring} &
\texttt{ringinfo.NumAtomRings(idx)} &
Number of rings that contain the atom (ring membership count). \\

\texttt{implicit\_valence} &
\texttt{atom.GetImplicitValence()} &
Implicit valence inferred by RDKit. \\

\texttt{formal\_charge} &
\texttt{atom.GetFormalCharge()} &
Formal charge. \\

\texttt{numH} &
\texttt{atom.GetTotalNumHs()} &
Total number of attached hydrogens. \\

\texttt{hybridization} &
\texttt{str(atom.GetHybridization())} &
Hybridization state (e.g., SP, SP2, SP3). \\

\texttt{is\_aromatic} &
\texttt{atom.GetIsAromatic()} &
Aromaticity flag. \\

\texttt{is\_in\_ring5} &
\texttt{ringinfo.IsAtomInRingOfSize(idx, 5)} &
Boolean indicator for membership in any 5-membered ring. \\

\texttt{is\_in\_ring6} &
\texttt{ringinfo.IsAtomInRingOfSize(idx, 6)} &
Boolean indicator for membership in any 6-membered ring. \\
\bottomrule
\end{tabular}
\end{table}

\subsubsection{Bond features (\texttt{featurize\_bond})}
For each \texttt{RDKit} bond object \texttt{bond} connecting atoms $i$ and $j$, we extract the bond type and whether the bond is conjugated (Table~\ref{tab:rdkit_bond_features}). Each attribute is mapped to an integer index against \texttt{bond\_features\_list}.

\begin{table}[t]
\centering
\small
\setlength{\tabcolsep}{7pt}
\renewcommand{\arraystretch}{1.15}
\caption{RDKit-derived bond features used in \texttt{featurize\_bond}.}
\label{tab:rdkit_bond_features}
\begin{tabular}{p{0.26\linewidth} p{0.40\linewidth} p{0.26\linewidth}}
\toprule
\textbf{Feature key} & \textbf{RDKit source} & \textbf{Description} \\
\midrule
\texttt{bond\_type} &
\texttt{str(bond.GetBondType())} &
Bond order/type (e.g., SINGLE, DOUBLE, TRIPLE, AROMATIC; otherwise mapped to ``misc''). \\

\texttt{is\_conjugated} &
\texttt{bond.GetIsConjugated()} &
Boolean flag indicating whether the bond is conjugated. \\
\bottomrule
\end{tabular}
\end{table}

\end{appendices}

\end{document}